\documentclass[11pt,a4paper]{article}
\usepackage{jheppub}

\usepackage{graphicx}
\usepackage{amsmath,amssymb,amsfonts}
\usepackage{verbatim}   
\usepackage{subfigure}  
\usepackage{acronym}
\usepackage{hyperref}
\usepackage{mathrsfs}
\usepackage{multirow}
\usepackage{slashed}
\usepackage{url}

\usepackage{amsmath}
\usepackage{amsfonts}
\usepackage{amssymb}

\usepackage{ulem}
\newcommand\POWHEG{{\tt POWHEG}}
\newcommand\BOX{{\tt POWHEG~BOX}}
\newcommand\MCatNLO{{\tt MC@NLO}}
\newcommand\SatM{{\tt S@M}}

\title{QCD effects in mono-jet searches for dark matter}
 
\preprint{\null\hfill {OUTP-13-20P}\\ \null\hfill {NSF-KITP-13-241}}

\author[a,b]{Ulrich Haisch,}
\author[a,b]{Felix Kahlhoefer}
\author[a]{and Emanuele Re}
\affiliation[a]{Rudolf Peierls Centre for Theoretical Physics, University of Oxford, \\ 1 Keble Road, Oxford OX1 3NP, United Kingdom}
\affiliation[b]{KITP, University of California Santa Barbara,\\ Santa Barbara, CA 93106-4030, U.S.A.}
\emailAdd{u.haisch1@physics.ox.ac.uk}
\emailAdd{felix.kahlhoefer@physics.ox.ac.uk}
\emailAdd{emanuele.re@physics.ox.ac.uk}

\abstract{LHC searches for missing transverse energy in association with a jet allow to place strong bounds on the interactions between dark matter and quarks. In this article, we present an extension of the POWHEG BOX capable of calculating the underlying cross sections at the next-to-leading order level. This approach enables us to consistently include the effects of parton showering and to apply realistic experimental cuts. We find significant differences from a fixed-order analysis that neglects parton showering effects. In particular, next-to-leading order corrections do not lead to a significant enhancement of the mono-jet cross section once a veto on additional jets is imposed. Nevertheless, these corrections reduce the theoretical uncertainties of the signal prediction and therefore improve the reliability of the derived bounds. We present our results in terms of simple rescaling factors, which can be directly applied to existing experimental analyses and discuss the impact of changing experimental cuts.}


\date{\today}

\begin{document}

\maketitle

\section{Introduction}
\label{sec:intro}

The clear evidence for the gravitational effects of dark matter (DM) clashes with our complete lack of knowledge about its nature and origin. This tension has led to a large number of proposed candidates for the particles that could constitute DM. Among the best motivated and most popular ones are weakly interacting massive particles, which can account for the observed DM abundance if they have a mass $m_\chi$ of $\mathcal{O}(100\,\text{GeV})$. These particles have been searched for in so-called direct detection experiments employing shielded underground detectors~---~so far without success (see~e.g.~\cite{Ahmed:2009zw, Aprile:2012nq}). However, as a by-product of these searches there have been various experimental hints for lighter DM particles with $m_\chi \lesssim 10\,\text{GeV}$~\cite{Bernabei:2010mq, Aalseth:2011wp, Angloher:2011uu, Agnese:2013rvf} leading to an increasing amount of theoretical interest in this mass range. Intriguingly, stable particles with a mass of $\mathcal{O}(5 \,\text{GeV})$ can account for  DM if they carry the same matter-antimatter asymmetry as baryons.

In the light of these recent experimental claims, it is of great importance to find complementary techniques to probe the low-mass region and either to confirm the DM hypothesis or to constrain the parameter space. This goal is difficult to achieve with direct detection experiments since the typical energy transfer in the scattering of such particles is small compared to the experimental energy thresholds and the interpretation of experimental results are affected by astrophysical uncertainties. Light DM particles produced at the LHC, on the other hand, can carry large amounts of momentum and therefore give clean signals that are independent of such uncertainties. 

Direct detection experiments are furthermore complicated by the fact that in the non-relativistic limit DM particles scatter coherently off the entire nucleus. For certain interactions, this coherence can lead to a large enhancement of the scattering cross section, but this enhancement depends on the ratio of the DM couplings to protons and neutrons and is altogether absent for spin-dependent interactions. These theoretical uncertainties make direct detection experiments difficult to interpret in terms of the underlying model. In contrast, at the LHC one can directly probe the interactions between DM particles and individual standard model (SM) particles.

The minimal experimental signature of DM production at the LHC would be an excess of events with a single jet or photon in association with large amounts of missing transverse energy~($E_{T,{\rm miss}}$). While these channels provide insufficient information to determine the mass of the DM particles, they can be used to constrain its scattering cross section in a  very general way. Such searches have been carried out at CDF~\cite{CDF-monojet}, CMS~\cite{Chatrchyan:2012me, CMS:rwa, Chatrchyan:2012tea} and ATLAS~\cite{ATLAS:2012ky, ATLAS:2012zim, Aad:2012fw} and lead to bounds which are comparable with or even superior to the ones obtained from direct detection experiments for low-mass DM~\cite{Goodman:2010yf,Bai:2010hh, Rajaraman:2011wf,Goodman:2010ku,Fox:2011fx, Fox:2011pm, Zhou:2013fla}.

Unfortunately, the backgrounds in mono-jet and mono-photon searches are large and the transverse momentum ($p_T$) spectrum of the signal is essentially featureless although it is slightly harder than that of the background. Consequently, the current sensitivity is already limited by systematical uncertainties. A combination of experimental and theoretical efforts will therefore be needed  to improve the reach of future searches. From the theoretical side, this will require calculating both background and signal predictions to greater accuracy. 

The dominant backgrounds, resulting from the production of SM vector bosons in association with a jet or a photon, have been known to next-to-leading order (NLO) for a long time~\cite{Giele:1991vf,Baur:1997kz}. More recently, attention has been paid to the importance of loop corrections for the signal process,~i.e.~the production of DM pairs plus a jet or photon~\cite{Haisch:2012kf,Fox:2012ru}. In the article~\cite{Fox:2012ru} the parton-level cross sections for these signals have been calculated at NLO and implemented into MCFM~\cite{MCFM}. These corrections do not only reduce the factorisation and renormalisation scale dependencies and hence the theoretical uncertainty of the signal prediction, but also lead to an overall increase of the cross sections. As a result, the NLO bounds are found to be both stronger and more reliable than those obtained at leading order (LO). 

Ideally, the LHC collaborations should be able to use an NLO implementation of the expected DM signal in order to optimise their cuts in such a way that backgrounds are reduced and uncertainties are minimised. For this purpose, a parton-level implementation is insufficient, because a full event simulation including showering and hadronisation is required. This can be achieved using a NLOPS method,~i.e.~an approach that allows to match consistently an NLO computation with a parton shower~(PS). Two of these approaches, namely \POWHEG{}~\cite{Nason:2004rx, Frixione:2007vw} and \MCatNLO{}~\cite{Frixione:2002ik}, have been implemented in public codes, and these programs have by now become standard tools for LHC analyses.  In this article we present an extension of the \BOX{}~\cite{Alioli:2010xd}, which allows for the generation of mono-jet events from DM pair production at NLO including PS effects.

We find that the enhancement of the mono-jet cross section found in fixed-order NLO calculations is diminished once the PS effects are included and a jet veto is imposed. Taking these effects into account, we show that the NLOPS cross sections are comparable to those at LOPS for all types of interactions that we consider. The ratios of NLOPS to LOPS predictions presented in this work can readily be used to rescale the results of existing experimental analyses. The resulting bounds are not significantly stronger but more reliable, since the NLO corrections reduce the scale uncertainties of the signal prediction.

The structure of this paper is as follows: in section~\ref{sec:effective} we introduce the general formalism and define the effective interactions which we consider subsequently. The results of our calculations, in particular a comparison of LO and NLO cross sections and distributions without and with PS effects, are presented in section~\ref{sec:LONLO}. Finally, in section~\ref{sec:bounds} we apply our results to the most recent mono-jet search performed by the CMS collaboration in order to obtain constraints on the DM scattering cross section. In appendix~\ref{sec:powheg} we provide details on our implementation in the \BOX{}.  A concise description of how to use the \BOX{} code is given in appendix~\ref{sec:manual}.

\section{Dark matter interactions}
\label{sec:effective}

In this work we are interested in DM pair production from quark or gluon initial states. We will restrict our discussion to the case where the production proceeds via the exchange of either a spin-$0$  or a spin-$1$ $s$-channel mediator. We consider the following interactions between DM and SM fields involving a scalar ($S$) or pseudo-scalar ($P$) mediator: 
\begin{equation} \label{eq:Lphi}
\begin{split}
& \mathcal{L}_S   = g_\chi^S \left(  \bar \chi   \chi  \right ) S + \sum_{q}  g_q^S \left(  \bar q q  \right ) S + \frac{\alpha_s}{\bar \Lambda}\, g_G^S \,    G_{\mu \nu}^a   G^{a, \mu \nu} S  \,, \\[1mm]
& \hspace{10mm} \mathcal{L}_P   =  i g_\chi^P \left(  \bar \chi  \gamma_5 \chi  \right ) P + \sum_{q} i g_q^P \left(  \bar q   \gamma_5 q  \right )P \,.
\end{split}
\end{equation}
The interactions including a vector ($V$) mediator take the form 
\begin{equation} \label{eq:LV}
\mathcal{L}_V  = \left ( \bar \chi \hspace{0.5mm} \gamma_\mu \left[ g^V_\chi \, + g^A_\chi \gamma_5 \right] \chi \right) V^\mu  + \sum_{q} \left( \bar q \hspace{0.5mm}  \gamma_\mu \left [  g^V_q  + g^A_q \gamma_5 \right] q  \right) V^\mu \,.
\end{equation}
Here we have assumed that the DM particle $\chi$ is a Dirac fermion, but extending our discussion to Majorana DM is straightforward.\footnote{For the operators ${\cal O}_A$, ${\cal O}_S$, ${\cal O}_P$ and ${\cal O}_G$  defined in (\ref{eq:QVQA}) to  (\ref{eq:QGQGt}) the predicted cross sections for Majorana DM are larger by a factor of 2. The analog of ${\cal O}_V$ with Majorana fermions is an evanescent operator.} The effective coupling of $S$  to gluons in (\ref{eq:Lphi}) can arise through loops of~e.g.~very heavy coloured fermions that couple to the mediator in a similar way as SM quarks. We parameterise this loop suppression by the factor $\alpha_s/\bar \Lambda$ and assume that the scale $\bar \Lambda$ is sufficiently high, that it cannot be probed directly at the LHC.

If the mediator mass $M$ is large  compared to the invariant mass of the DM pair, we can describe DM pair production with an effective field theory (EFT). Integrating out the vector mediator gives rise to the vector and the axial-vector operators
\begin{equation} \label{eq:QVQA}
{\cal O}_V = \frac{1}{\Lambda^2} \left (\bar q \gamma_\mu q \right ) \left ( \bar \chi \gamma^\mu \chi \right )  \,, \qquad 
{\cal O}_A = \frac{1}{\Lambda^2} \left (\bar q \gamma_\mu \gamma _5 q \right )  \left ( \bar \chi \gamma^\mu \gamma_5 \chi \right ) \,,
\end{equation}
as well as two parity-violating operators from cross terms, which we do not consider further. We also ignore composite operators consisting of two SM currents, since interactions of this type are strongly constrained by di-jet searches \cite{Dreiner:2013vla}. The suppression scale $\Lambda$ is introduced in such a way as to make the effective interactions dimensionless. It should always be clear from the context which particular $\Lambda$ and operator we are studying, so we suppress the indices $V,A$ for $\Lambda$.

The scalar and pseudo-scalar interactions introduced in (\ref{eq:Lphi}) lead to the following two effective DM-quark interactions
\begin{equation} \label{eq:QSQP}
{\cal O}_S = \frac{m_q}{\Lambda^3} \left (\bar q q \right ) \left ( \bar \chi  \chi \right ) \,, \qquad 
{\cal O}_P = \frac{m_q}{\Lambda^3} \left (\bar q  \gamma _5 q \right ) \left ( \bar \chi \gamma_5 \chi \right )  \,,
\end{equation}
as well as  to the gluonic operator
\begin{equation} \label{eq:QGQGt}
{\cal O}_G = \frac{\alpha_s}{\Lambda^3} \, G_{\mu \nu}^a G^{a, \mu \nu} \left ( \bar \chi  \chi \right )  \,.
\end{equation}
Notice that in (\ref{eq:QSQP}) we have assumed that $S$ and $P$ couple to quarks proportional to their mass, i.e.~$g_q^{S,P} = g^{S,P} \, m_q / \bar \Lambda$, motivated by the hypothesis of minimal flavour violation, which curbs the size of dangerous flavour-changing neutral current processes~\cite{Batell:2011tc}. Note that, if DM couples to top quarks, the two operators ${\cal O}_{S,P}$ will induce large DM-gluon interactions via top-quark loops. These interactions have been discussed in detail in~\cite{Haisch:2012kf, Fox:2012ru}, so we restrict our attention to the light flavours $q = u,d,s,c,b$ here (see also~\cite{ Haisch:2013uaa,Lin:2013sca} for constraints on effective interactions between DM and top quarks).

While we have derived the effective operators above by integrating out an $s$-channel mediator, they can in principle arise from a wide range of different ultraviolet (UV) completions. Moreover, these effective operators have the advantage that they can be immediately applied to different processes, such as the scattering of DM particles on nucleons in direct detection experiments. To give an example, in the case of the vector operator the induced DM-proton scattering cross section takes the form
\begin{equation}
\sigma_p = \frac{f_p^2}{\pi}\frac{m_{\rm red}^2}{\Lambda^4} \,, 
\end{equation}
where $f_p = 3$ is the effective DM-proton coupling and $m_{\rm red} = m_p m_\chi/(m_p + m_\chi)$ is the reduced mass of the DM-proton system. Similar expressions can be derived for the other effective operators (see e.g.~\cite{Beltran:2008xg,Agrawal:2010fh,Rajaraman:2011wf,MarchRussell:2012hi}). We will therefore use the effective operators from above to interpret experimental data. The interested reader is referred to~\cite{Fox:2011pm,Shoemaker:2011vi,Fox:2012ee,Busoni:2013lha,Profumo:2013hqa, Buchmueller:2013dya} for discussions of the validity of the EFT and to~\cite{Frandsen:2012rk} for how constraints on DM parameters can still be obtained in the case that the mediator is too light to be integrated out.

\section{Impact of NLO corrections and showering}
\label{sec:LONLO}

In this section we present our results for the  fixed-order parton-level predictions at LO and NLO and compare them with those after showering and hadronisation.  We consider jet~+~$E_{T,{\rm miss}}$~production at the LHC with $\sqrt{s}=8$~TeV centre-of-mass~(CM) energy. Unless otherwise stated, we have performed all simulations using the EFT approach introduced above, setting $\Lambda=500$~GeV.   Our  LO and NLO predictions are obtained using the MSTW2008~LO and NLO parton distribution functions (PDFs)~\cite{Martin:2009iq} and the corresponding  reference value for the strong coupling constant. We find the scale $\mu$ which determines $\alpha_s(\mu)$ dynamically, i.e.~we define $\mu  = \xi \hspace{0.25mm} H_T/2 = \mu_R = \mu_F$ and evaluate it on an event-by-event basis.  Here 
\begin{equation} \label{eq:HT}
H_T =   \sqrt{m_{ \bar \chi \chi}^2 + p_{T,j_1}^2} + p_{T,j_1} \, ,
\end{equation}
with $m_{ \bar \chi \chi}$ denoting the invariant mass of the DM pair and $p_{T,j_1}$  the transverse momentum of the hardest jet $j_1$. To assess the theoretical errors in our analysis, we study the ambiguities related to a variation of the renormalisation ($\mu_R$) and factorisation ($\mu_F$) scale by varying $\xi$ in the range $[1/2 , 2]$. As will see below, our scale choice has the advantage that the size of NLO corrections is largely independent of the DM mass $m_\chi$.

\begin{table}
\begin{center}
\begin{tabular}[!t]{|c||c|}
\hline
CMS & ATLAS \\
\hline \hline
 $|\eta_j|<4.5$, $p_{T,j}>30 \, {\rm GeV}$, $N_j \leq 2$ &   $|\eta_j|<4.5$, $p_{T,j}>30 \, {\rm GeV}$, $N_j \leq 2$ \\
 $\Delta\phi_{j_1,j_2}<2.5$ &  $\Delta\phi_{j_2,\vec{E}_{T,{\rm miss}}}>0.5$ \\
 $|\eta_{j_1}|<2.4$, $p_{T,j_1}>110 \, {\rm GeV}$,  $E_{T,{\rm miss}}> 350 \, {\rm GeV}$ & $|\eta_{j_1}|<2$, \, $p_{T,j_1}, E_{T,{\rm miss}}  > 350 \, {\rm GeV}$ \\
\hline 
\end{tabular}
\end{center}
  \caption{\label{tab:cuts} Event selection criteria applied in our analysis. See text for further explanations.}
\end{table}

In our analysis we adopt two sets of cuts  corresponding to the latest  CMS~\cite{CMS:rwa}  and ATLAS~\cite{ATLAS:2012zim}   mono-jet search summarised in table~\ref{tab:cuts}.  Both experiments reject events with more than two jets with  pseudo-rapidity below 4.5 and transverse momentum above $30 \, {\rm GeV}$~($N_j \leq 2$). We construct jets according to the anti-$k_t$ algorithm \cite{Cacciari:2005hq,Cacciari:2008gp}, as implemented in {\tt FastJet} \cite{Cacciari:2011ma}, using a radius parameter of $R=0.4$.\footnote{The CMS collaboration uses $R =0.5$, while ATLAS employs $R =0.4$ in their mono-jet searches. Here we adopt $R = 0.4$ for both searches to facilitate the comparison. Choosing $R = 0.5$ instead would increase the predicted cross sections by 3\% to 4\%, while the $K$ factors change by less than 1\%. The $K$ factors presented below can hence be used for both CMS and ATLAS.} In order to suppress QCD di-jet events, CMS puts an angular requirement on $\Delta\phi_{j_1,j_2}$, while ATLAS cuts on the azimuthal separation $\Delta\phi_{j_2,E_{T,{\rm miss}}}$ to reduce the background originating from the mis-measurement of  the transverse momentum of the second-leading jet $j_2$. The signal region is defined in the case of the CMS search by $|\eta_{j_1}|<2.4$, $p_{T,j_1}>110 \, {\rm GeV}$ and $E_{T,{\rm miss}}> 350 \, {\rm GeV}$, while ATLAS  imposes the cuts $|\eta_{j_1}|<2$ and $p_{T,j_1},  E_{T,{\rm miss}}  > 350 \, {\rm GeV}$. Clearly, apart from the leading-jet and $E_{T,{\rm miss}}$ requirements the event selection criteria in both analyses are quite similar. Nevertheless, we will see below that there are important differences between the two analyses concerning the impact of NLO and PS effects.

\subsection{Vector and axial-vector operators}
\label{sec:VA}

\subsubsection{CMS cuts}

\begin{figure}[t!]
\begin{center}
\includegraphics[width=0.5\columnwidth]{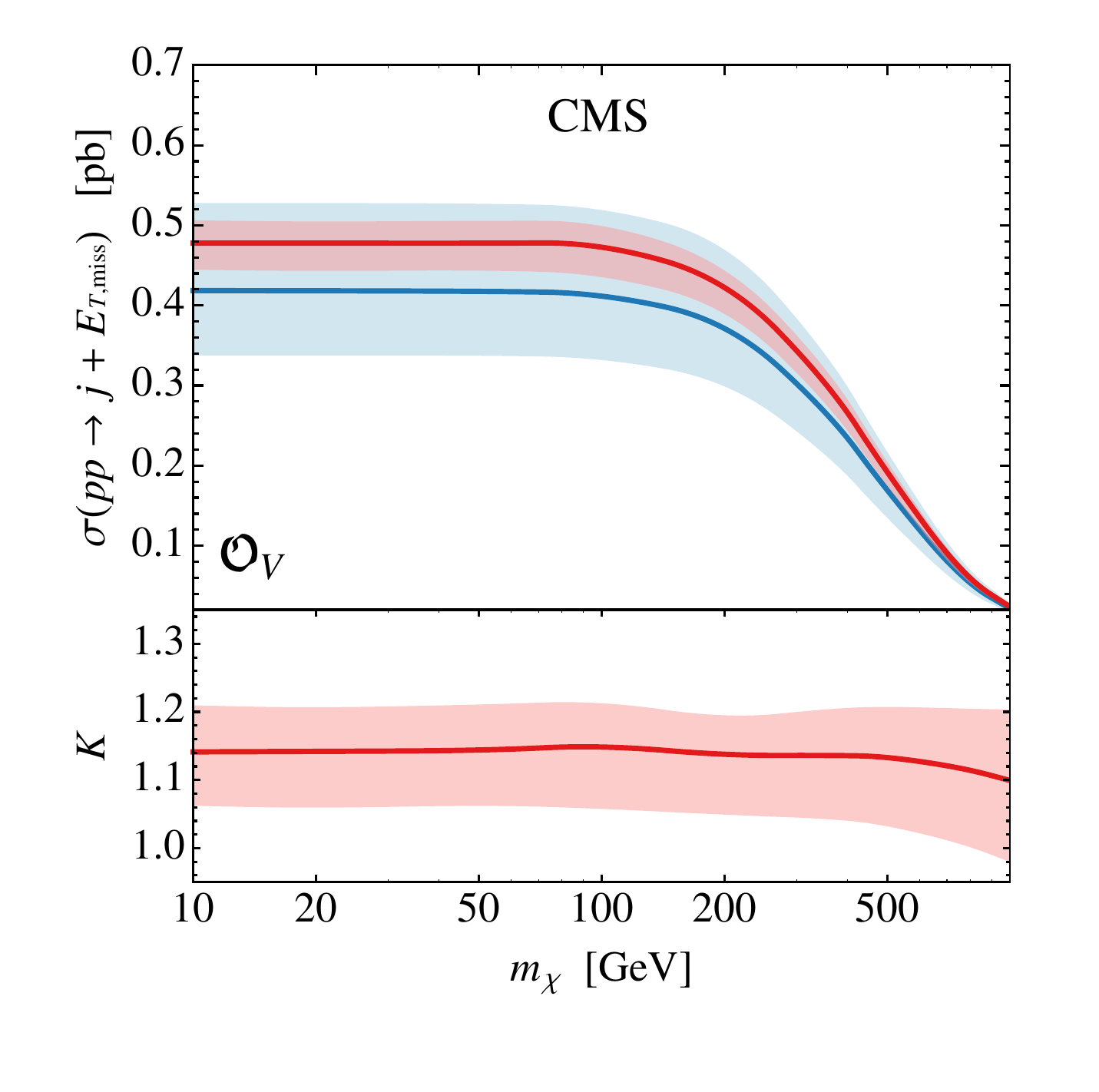}   \hspace{-4mm} 
\includegraphics[width=0.5\columnwidth]{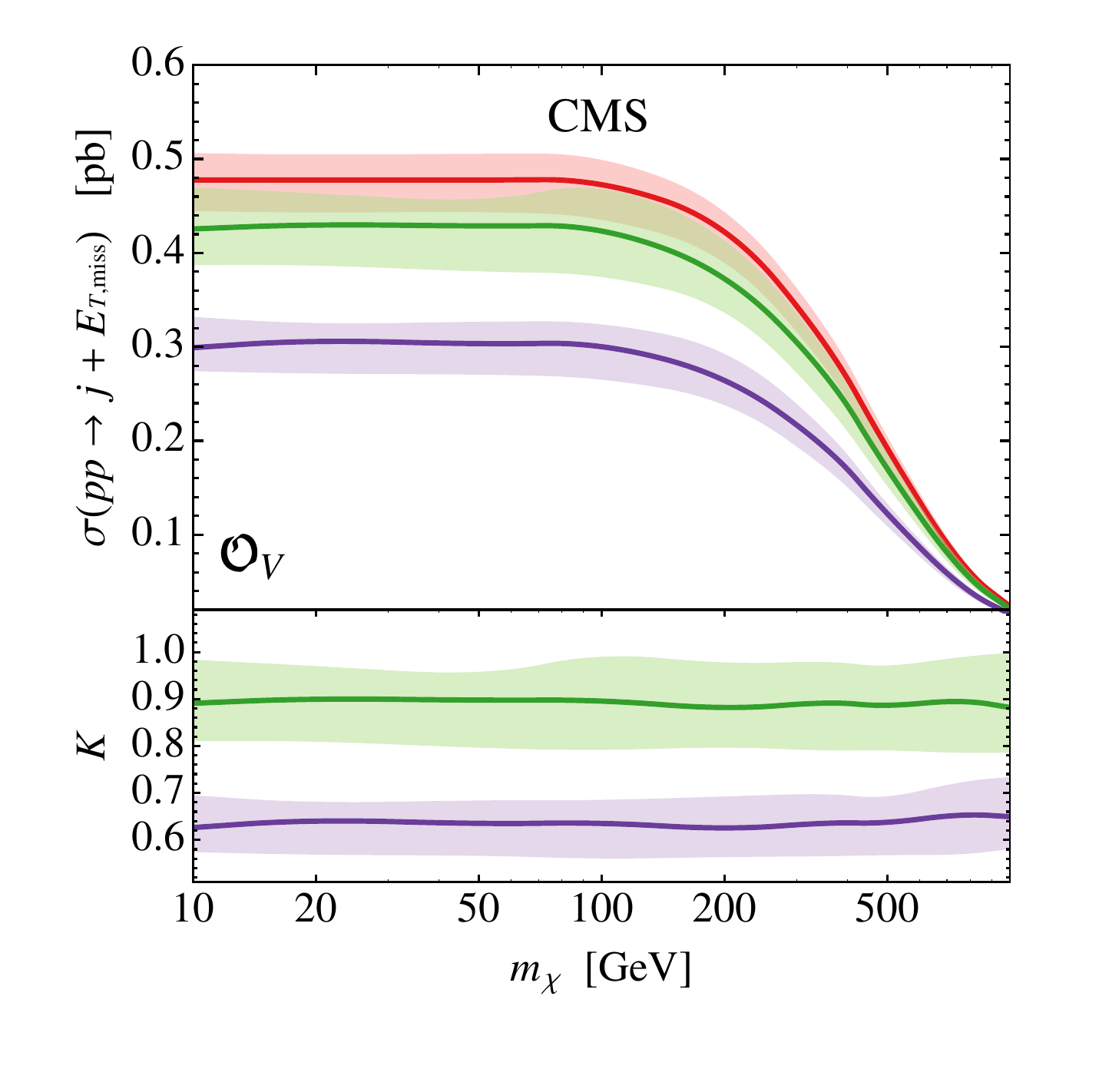}   
\end{center}
\vspace{-8mm}
\caption{\label{fig:CMSxsec} Left panel: LO (blue) and NLO (red) fixed-order results for the mono-jet cross section and the corresponding $K$ factor. Right panel: Fixed-order NLO result (red), the inclusive NLOPS  prediction (green) and the NLOPS result with jet veto (purple).  The shown predictions correspond to the vector operator ${\cal O}_V$ and the CMS  event selection criteria.}
\end{figure}

We begin our numerical analysis by considering the predictions for the mono-jet cross section obtained for the vector operator (\ref{eq:QVQA}) by employing the CMS cuts. Our results are given in figure~\ref{fig:CMSxsec}. The left panel shows the fixed-order predictions (i.e.~without PS effects) with the width of the coloured bands reflecting the associated scale uncertainties. One observes that the scale dependencies of the LO prediction amount to around $^{+25\%}_{-20\%}$ and are reduced to about $^{+9\%}_{-6\%}$ after including NLO corrections. The $K$ factor, defined as 
\begin{equation} \label{eq:Kdef}
K = \frac{\sigma (pp \to j + E_{T,{\rm miss}})_{\rm NLO}^{\xi = [1/2,2]} }{\sigma (pp \to j + E_{T,{\rm miss}})_{\rm LO}^{\xi=1}} \,,
\end{equation}
is roughly 1.1, meaning that NLO effects slightly enhance the mono-jet cross section with respect to the LO result. Moreover, we find that the $K$ factor is almost independent of the DM mass. This stability is related to our choice of scales (\ref{eq:HT}) and should be contrasted with the results in \cite{Fox:2012ru} that employ $\mu = m_{ \bar \chi \chi} = \mu_R = \mu_F$ as the central scale. Compared to our scale setting the latter  choice tends to underestimate the LO cross sections for heavy DM particles, which leads to an artificial rise of the $K$ factor.  

In the right panel of figure~\ref{fig:CMSxsec} we compare the fixed-order NLO prediction with the NLOPS results obtained in the \BOX{}  framework using PYTHIA 6.4~\cite{Sjostrand:2006za} for showering and hadronisation. The shown $K$ factors are defined relative to the fixed-order NLO prediction in analogy to (\ref{eq:Kdef}). To better illustrate the effects of the PS we depict results for two different sets of cuts: the green curve and band correspond to an inclusive jet~+~$E_{T,{\rm miss}}$ search allowing for an arbitrary number $N_j$ of jets, while the purple curve and band correspond to the actual CMS analysis imposing a jet veto $N_j \leq 2$~(cf.~table~\ref{tab:cuts}). We observe that for a (hypothetical) inclusive mono-jet search the effects of showering and hadronisation are small, amounting to relative shifts in the range of $^{-2\%}_{-20\%}$.  This finding confirms the results  of~\cite{Bai:2010hh,Fox:2011pm}, extending them to the NLO level.  

In realistic jet + $E_{T,{\rm miss}}$ searches, however, the impact of showering and hadronisation is not  small. Once the number of jets is restricted to two or less, the NLOPS  cross section is visibly below the NLO prediction. Numerically, we find relative shifts  of~$^{-30\%}_{-45\%}$. The physical origin of the observed suppression of the cross section is clear:  events with one jet (two jets) that pass the cuts before showering will at the end be rejected, if soft QCD radiation associated to the PS is able to generate two additional jets~(one additional jet) with  $|\eta_j|<4.5$ and $p_{T,j}>30 \, {\rm GeV}$. The probability for this to happen is non-negligible (considering the large CM energy), and always leads to a drop in the number of  accepted events. 

\subsubsection{ATLAS cuts}

\begin{figure}[t!]
\begin{center}
\includegraphics[width=0.5\columnwidth]{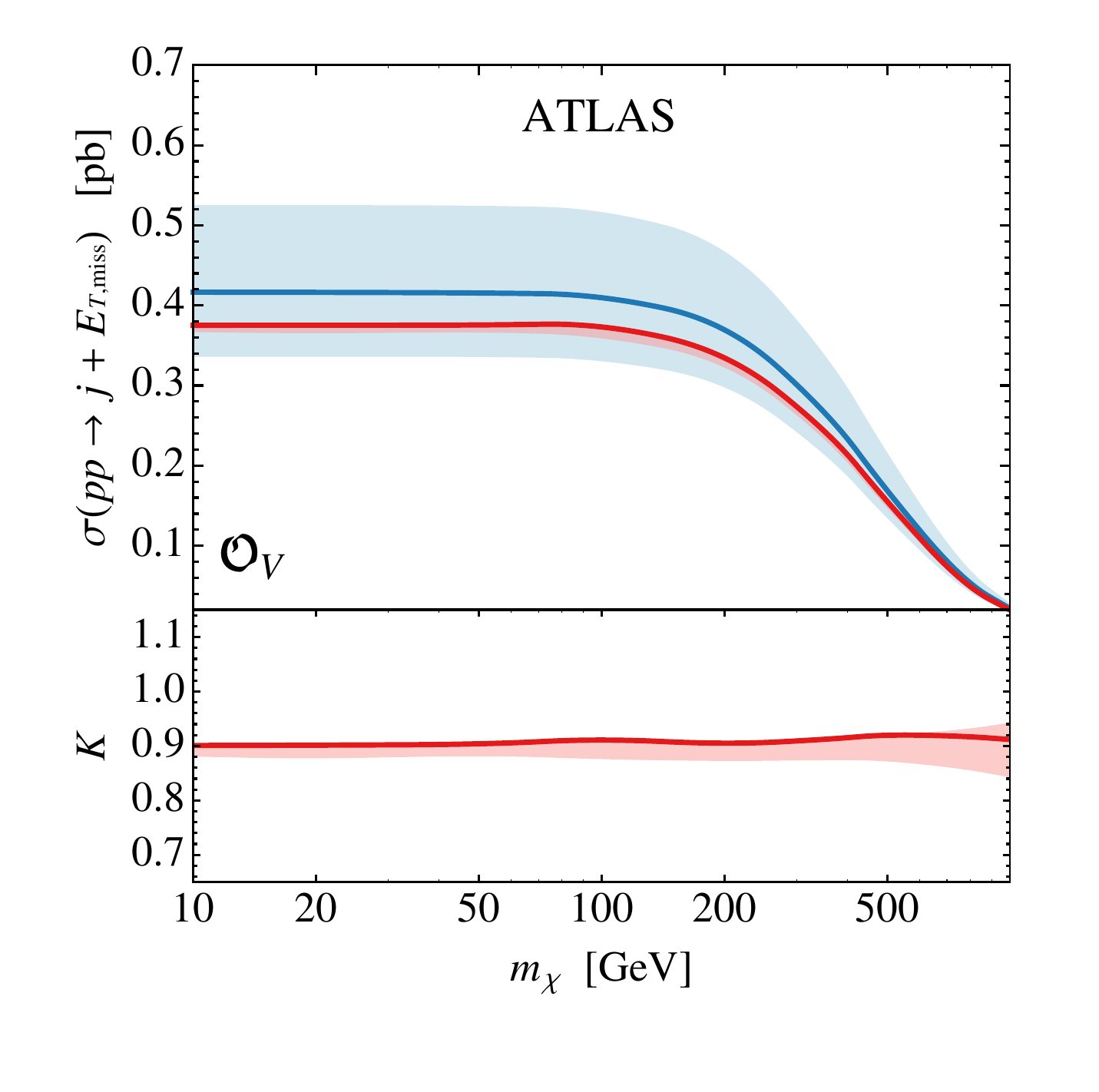}   \hspace{-4mm}
\includegraphics[width=0.5\columnwidth]{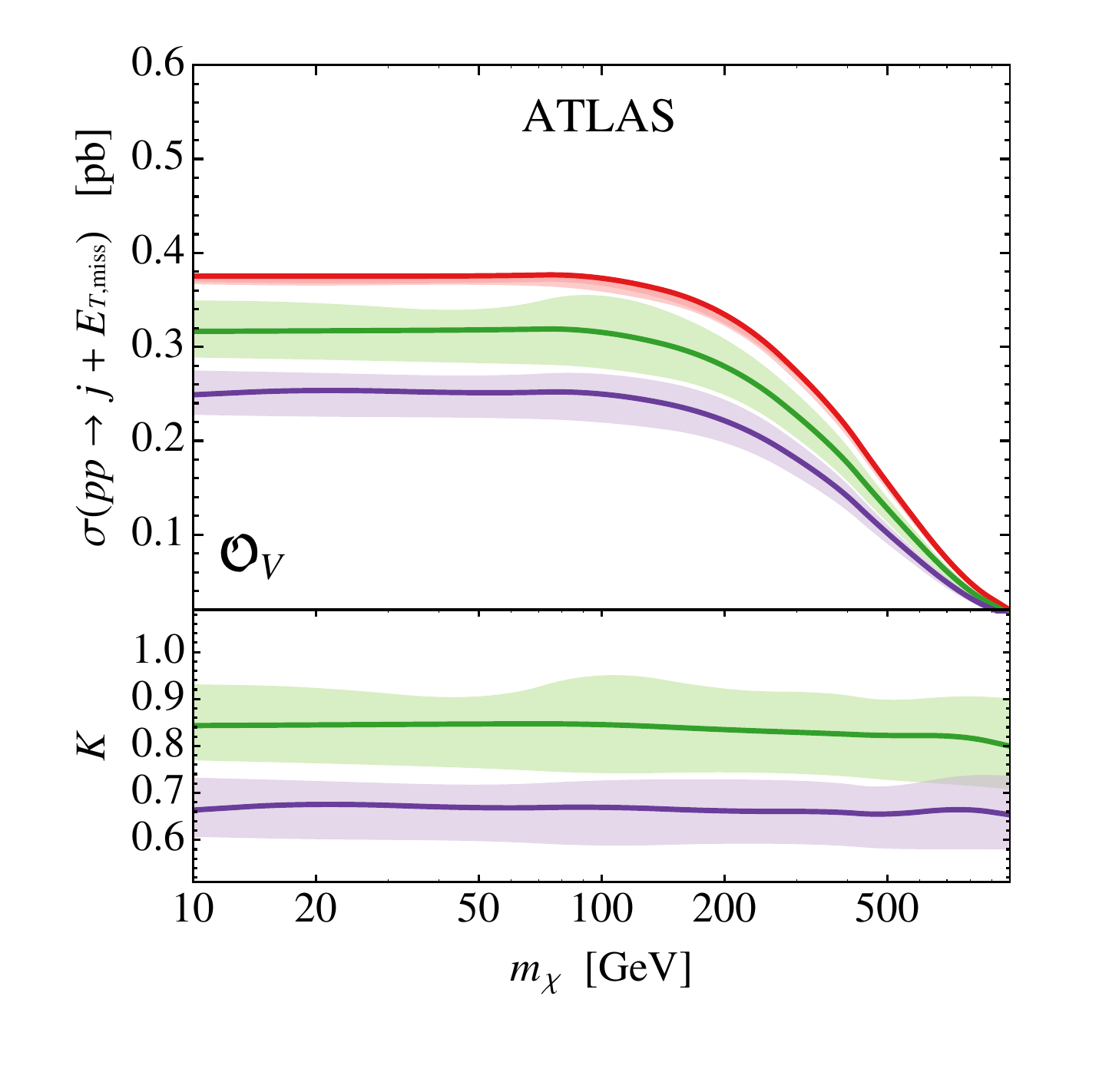}   
\end{center}
\vspace{-8mm}
\caption{\label{fig:ATLASxsec}  Comparison of  the predictions for the mono-jet cross section  associated to ${\cal O}_V$ applying the ATLAS  cuts. The same colour coding as in figure~\ref{fig:CMSxsec} is used.}
\end{figure}

We now turn our attention to the results obtained with the ATLAS cuts. The corresponding predictions are presented in the two panels of figure~\ref{fig:ATLASxsec}. In this case the NLO fixed-order prediction for the mono-jet cross section is below the LO result leading to a $K$ factor of about 0.9. This decrease results from the general tendency of non-soft QCD corrections to reduce the $p_T$ of the leading jet, which can thereby drop below the requirement imposed by ATLAS. However, the large reduction of scale uncertainties from~$^{+25\%}_{-20\%}$ to $^{+1\%}_{-4\%}$ is clearly pathologic, and does not represent a reliable measure of the theoretical uncertainties inherent in the NLO calculation. To verify that the cancellation of scale uncertainties is in large parts accidental, we have studied the dependence of the NLO cross section on the radius parameter $R$ and the choice of scale setting. We find that for jet radii larger than our reference value $R = 0.4$ the resulting scale ambiguities are more pronounced, and that scale choices different from (\ref{eq:HT}) also lead to bigger theoretical errors.

A related issue is that under certain circumstances phase-space cuts can lead to ill-behaved cross sections. For instance, in the case of di-jet rates it is well-known that fixed-order computations fail if symmetric transverse energy $E_T$ cuts are used on the two hardest jets~(see~e.g.~\cite{Frixione:1997ks}). The breakdown of the NLO prediction occurs in this case because the di-jet cross section becomes extremely sensitive to soft-gluon emission. The associated large logarithms then have to be resummed to all orders to obtain a meaningful result~\cite{Banfi:2003jj}. In the case of the jet + $E_{T,{\rm miss}}$ cross section such an infrared sensitivity does, however, not develop, since soft singularities are avoided because of the jet requirement $p_{T,j} > 30 \, {\rm GeV}$ (see~table~\ref{tab:cuts}). This means that the fixed-order NLO calculations of the DM signal do not break down, and no resummation of large logarithms is needed.

The particular impact of a symmetric cut on $p_{T,j_1} $ and $E_{T,{\rm miss}}$  is further illustrated by the right panel of figure~\ref{fig:ATLASxsec}, which compares the NLO results before and after showering and hadronisation. One observes that in this case the prediction for the inclusive NLOPS cross section (green curve and band) does not overlap with the (fixed-order) NLO  estimate. Shower and hadronisation effects are hence nominally larger for ATLAS than for CMS cuts, and amount to~$^{-8\%}_{-25\%}$. The impact of the veto $N_j \leq 2$ (purple curve and band), on the other hand, turns out to  give very similar results as for the CMS cuts, suppressing  the NLO cross section by $^{-27\%}_{-41\%}$.

\subsubsection{Comparison to LOPS}

\begin{figure}[t!]
\begin{center}
\includegraphics[width=0.5\columnwidth]{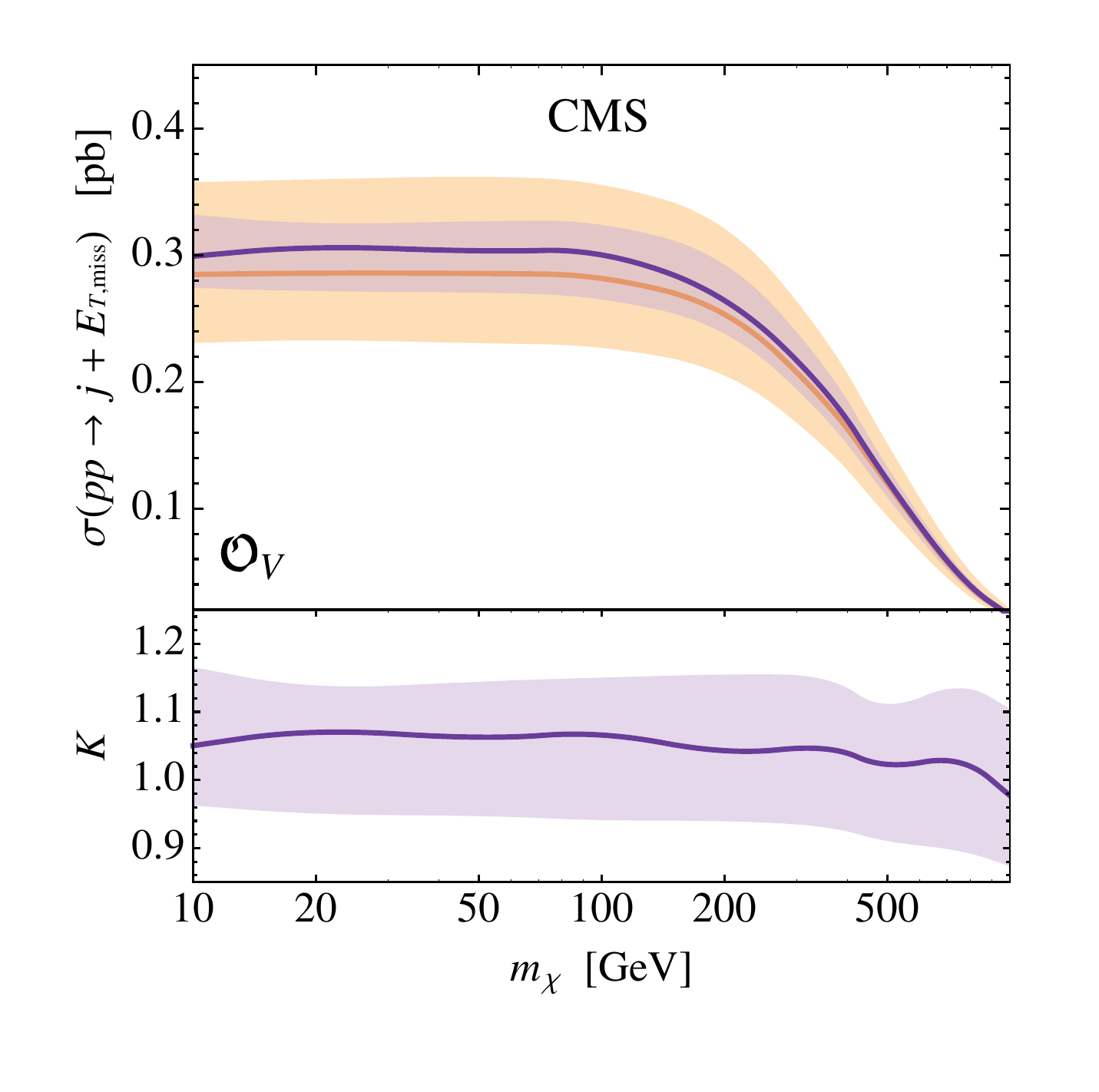}   \hspace{-4mm} 
\includegraphics[width=0.5\columnwidth]{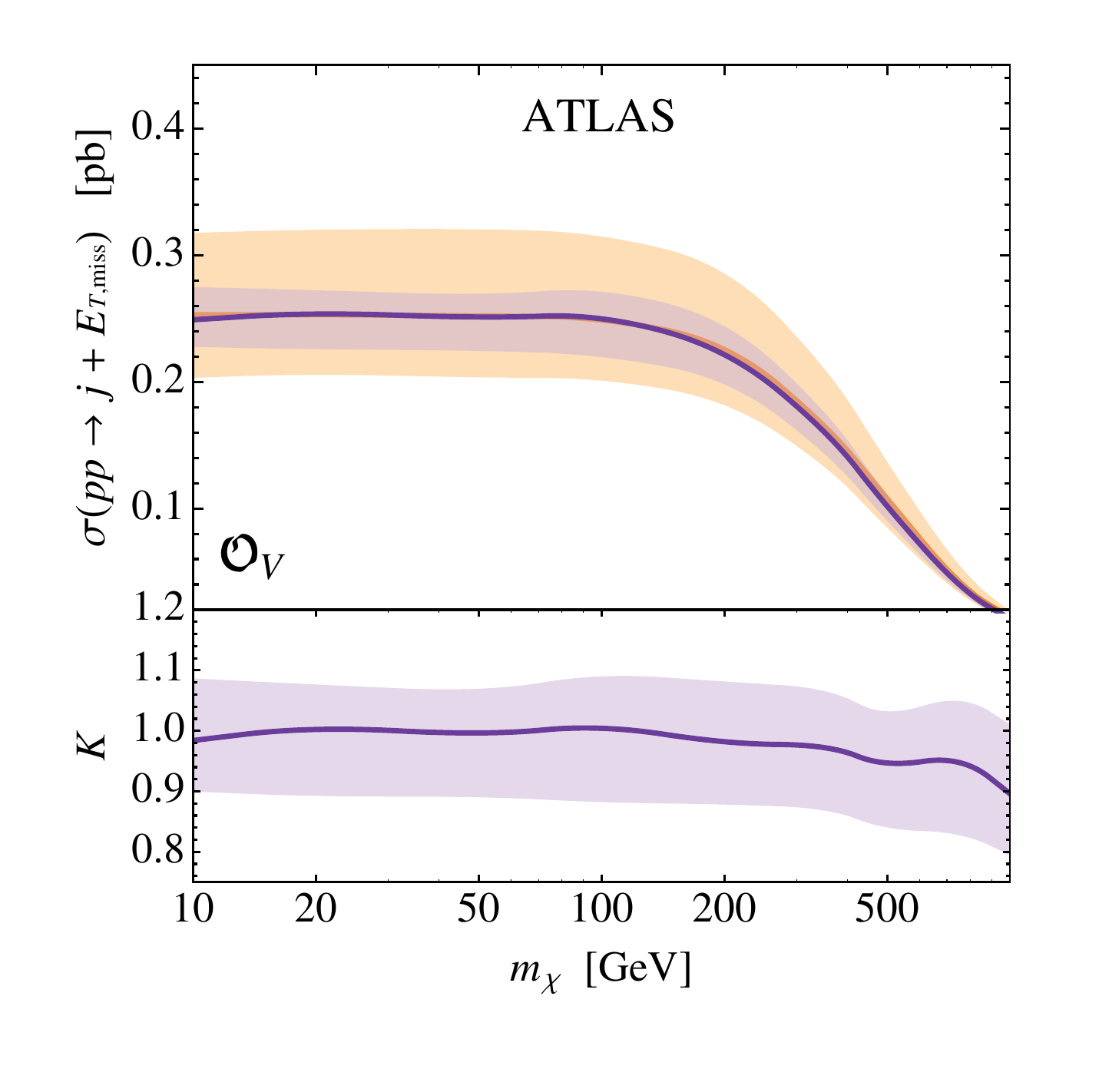}   
\end{center}
\vspace{-8mm}
\caption{\label{fig:realK}  NLOPS (purple  curve and band) and LOPS (orange curve and band) mono-jet cross sections with jet veto $N_j \leq 2$ and corresponding $K$ factor. The left (right) panel depicts the  results imposing the CMS (ATLAS) cuts. In both cases the insertion of the operator ${\cal O}_V$ is considered.}
\end{figure}

Since both ATLAS and CMS model the DM signal using an LOPS method,  we compare in figure~\ref{fig:realK} these results to the NLOPS predictions. The resulting $K$ factors can be used to promote the LOPS bounds on the suppression scale $\Lambda$ of the vector operator ${\cal O}_V$ derived in \cite{CMS:rwa, ATLAS:2012zim}  to the NLOPS level (taking into account that the bound on $\Lambda$ scale as $K^{1/4}$). We see that for both experimental settings, the inclusion of NLO effects leads to a notable reduction of theoretical uncertainties by   more than a factor of 2, and that the resulting $K$ factors are close to 1 and essentially flat with respect to $m_\chi$. Explicitly we find 
\begin{equation} \label{eq:KVCMSATLAS}
K_{\rm CMS}^V = 1.04^{+0.10}_{-0.11} \, , \qquad 
K_{\rm ATLAS}^V = 0.97^{+0.09}_{-0.11} \,,
\end{equation}
if the CMS and ATLAS selection criteria with jet veto are imposed.\footnote{Dropping the requirement $N_j \leq 2$ would result in $K$ factors that are larger by around $10\%$.} As for the fixed-order results we find $K_{\rm ATLAS}^V < K_{\rm CMS}^V$. The observed differences are however much smaller, since they are diluted by PS corrections. Consequently, applying fixed-order $K$ factors to rescale the bounds on $\Lambda$ obtained by LOPS calculations of the DM signal, would lead to limits that are too strong (weak) in the case of CMS (ATLAS). To reduce theoretical uncertainties on $\Lambda$ it is unavoidable to perform a NLOPS simulation that correctly takes into account the  selection cuts implemented in a given experimental analysis. 

\begin{figure}[t!]
\begin{center}
\includegraphics[width=0.29\columnwidth]{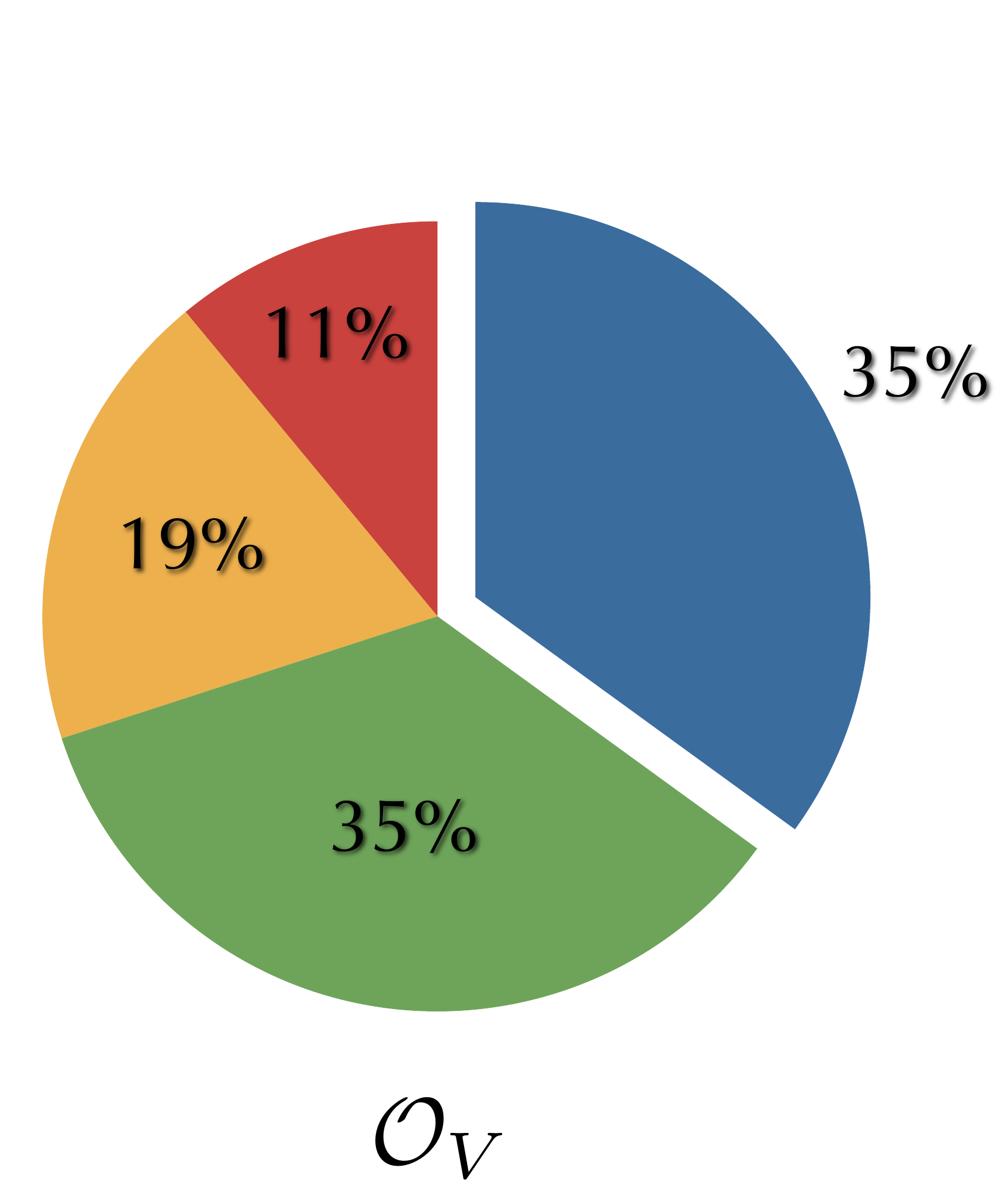} \qquad 
\includegraphics[width=0.25\columnwidth]{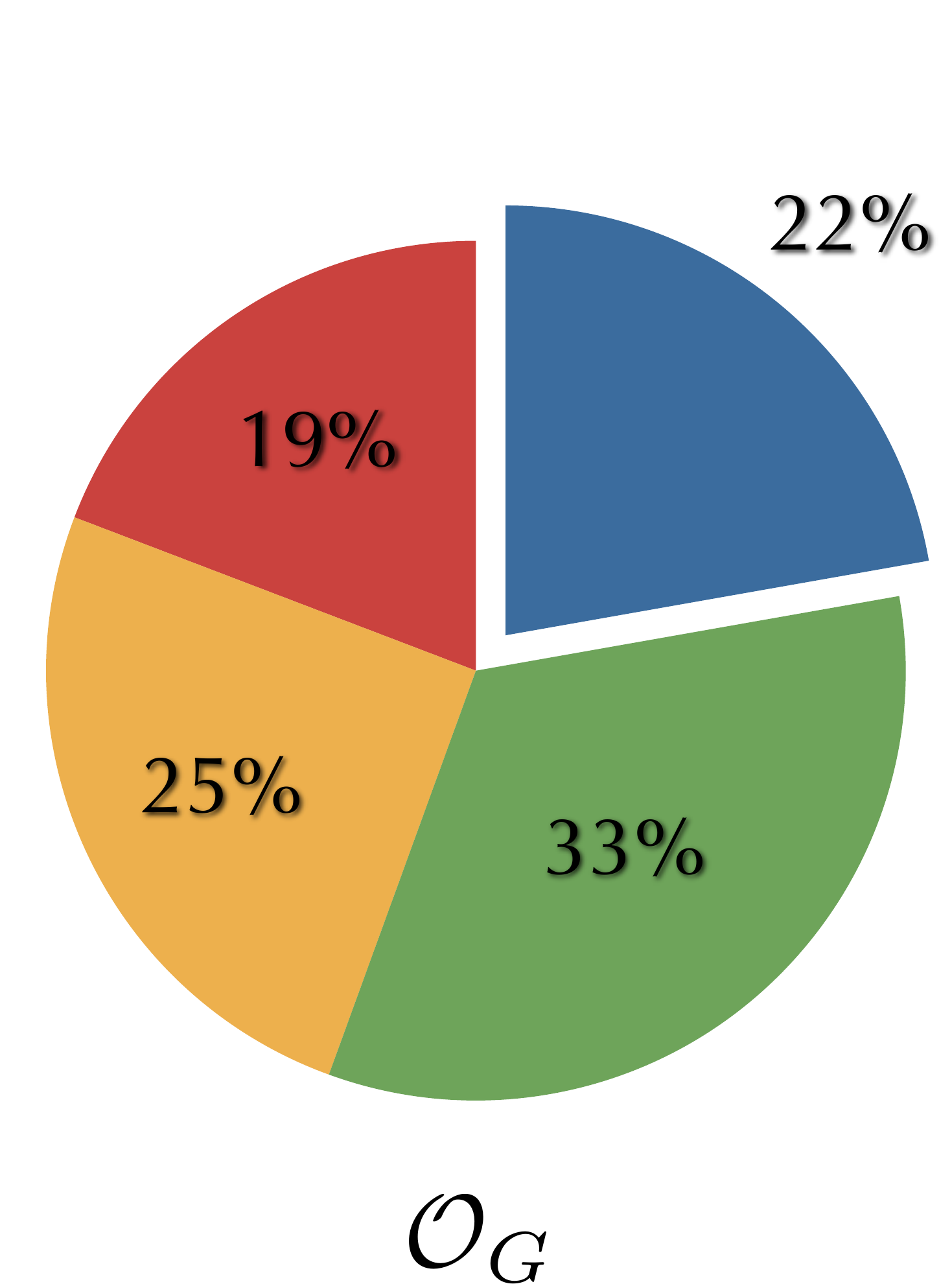} \qquad 
\includegraphics[width=0.28\columnwidth]{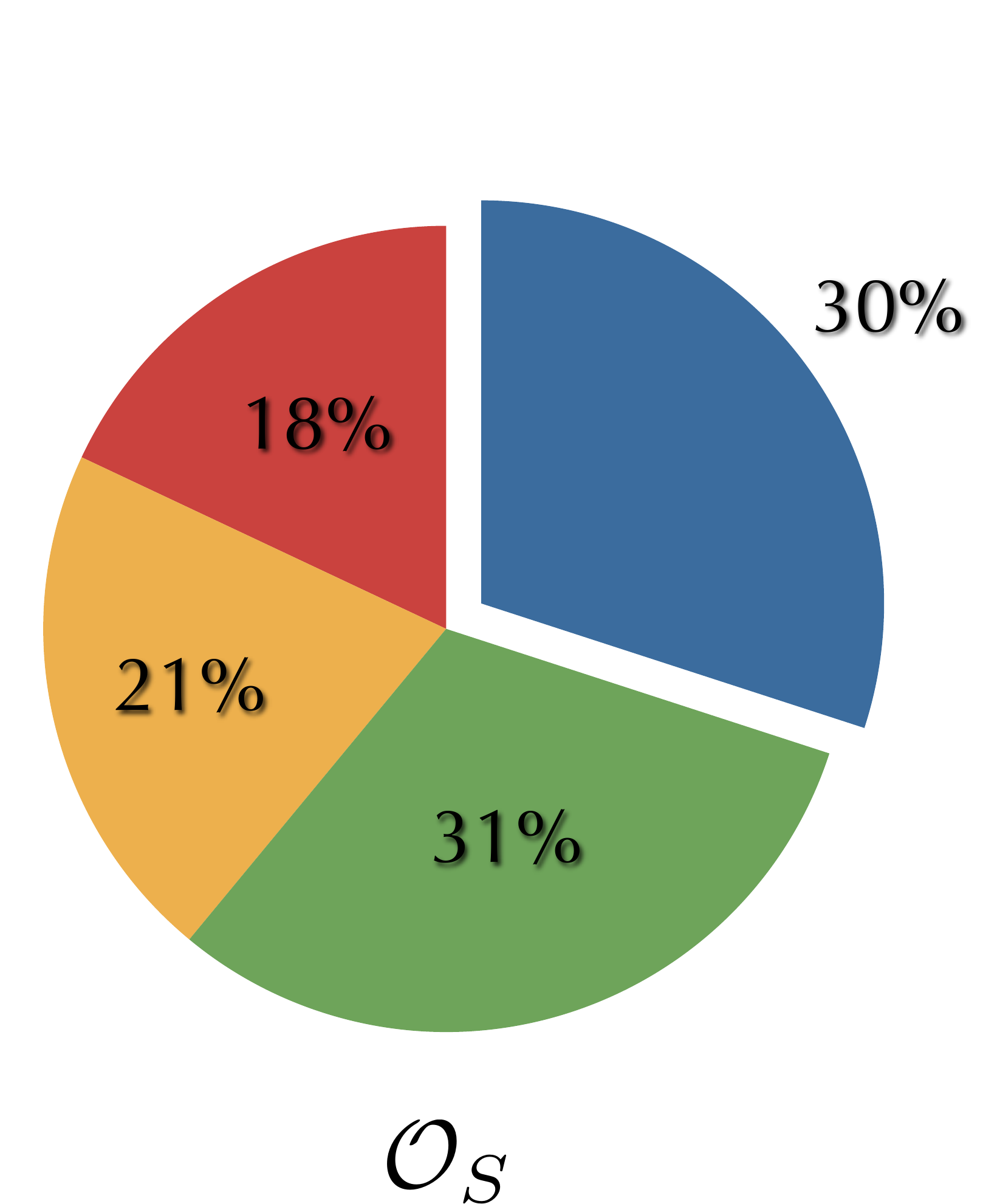} 
\end{center}
\vspace{-4mm}
\caption{\label{fig:pies} Fractions of events with $N_j =1$~(blue), $N_j =2$~(green),  $N_j =3$~(yellow) and  $N_j \geq 4$~(red) relative to the total jets + $E_{T, {\rm miss}}$ cross section. The results for ${\cal O}_V$, ${\cal O}_G$ and ${\cal O}_S$ are shown from left to right. All numbers correspond to CMS cuts and $m_\chi = 10 \, {\rm GeV}$.}
\end{figure}

In order to better understand the smallness of NLO effects implied by the results in~(\ref{eq:KVCMSATLAS}) we show on the left-hand side in figure~\ref{fig:pies} the fraction of $E_{T, {\rm miss}}$ events with exactly $1$ jet (blue pie piece), $2$ jets (green pie piece), $3$ jets (yellow pie piece) and more than 3 jets~(red pie piece) for the case of the vector operator ${\cal O}_V$. The given fractions correspond to CMS cuts and assume a DM mass of $10 \, {\rm GeV}$. One observes that~---~in spite of the name ``mono-jet search''~---~only 35\% of the events contain a single jet, while 65\%  of the total cross section is due to events with more than 1 jet.\footnote{For the ATLAS selection requirements these numbers change into 47\% and 53\%.} These numbers imply that most of the jets result either from  {\POWHEG} or from soft QCD that is modelled by the PS. Therefore the description of  events with $N_j \geq 2$ is at most LO accurate.  The loose jet cuts, i.e.~$|\eta_j|<4.5$ and  $p_T > 30 \, {\rm GeV}$, imposed on the non-leading jet in realistic jet~+~$E_{T, {\rm miss}}$ searches hence curb the impact  of the fixed-order NLO corrections that are included in our analysis  only for the $1$~jet + $E_{T,{\rm miss}}$ channel. The large importance of secondary jets hence reduces the impact  of the fixed-order NLO corrections for the $1$~jet + $E_{T,{\rm miss}}$ channel. We therefore expect that tighter cuts on secondary jets would  reduce theoretical uncertainties at the cost of diminishing the size of the expected signal. Quantifying the possible gain would require a dedicated experimental analysis including background estimates, which is beyond the scope of the present article.

\begin{figure}[t!]
\begin{center}
\includegraphics[width=0.5\columnwidth]{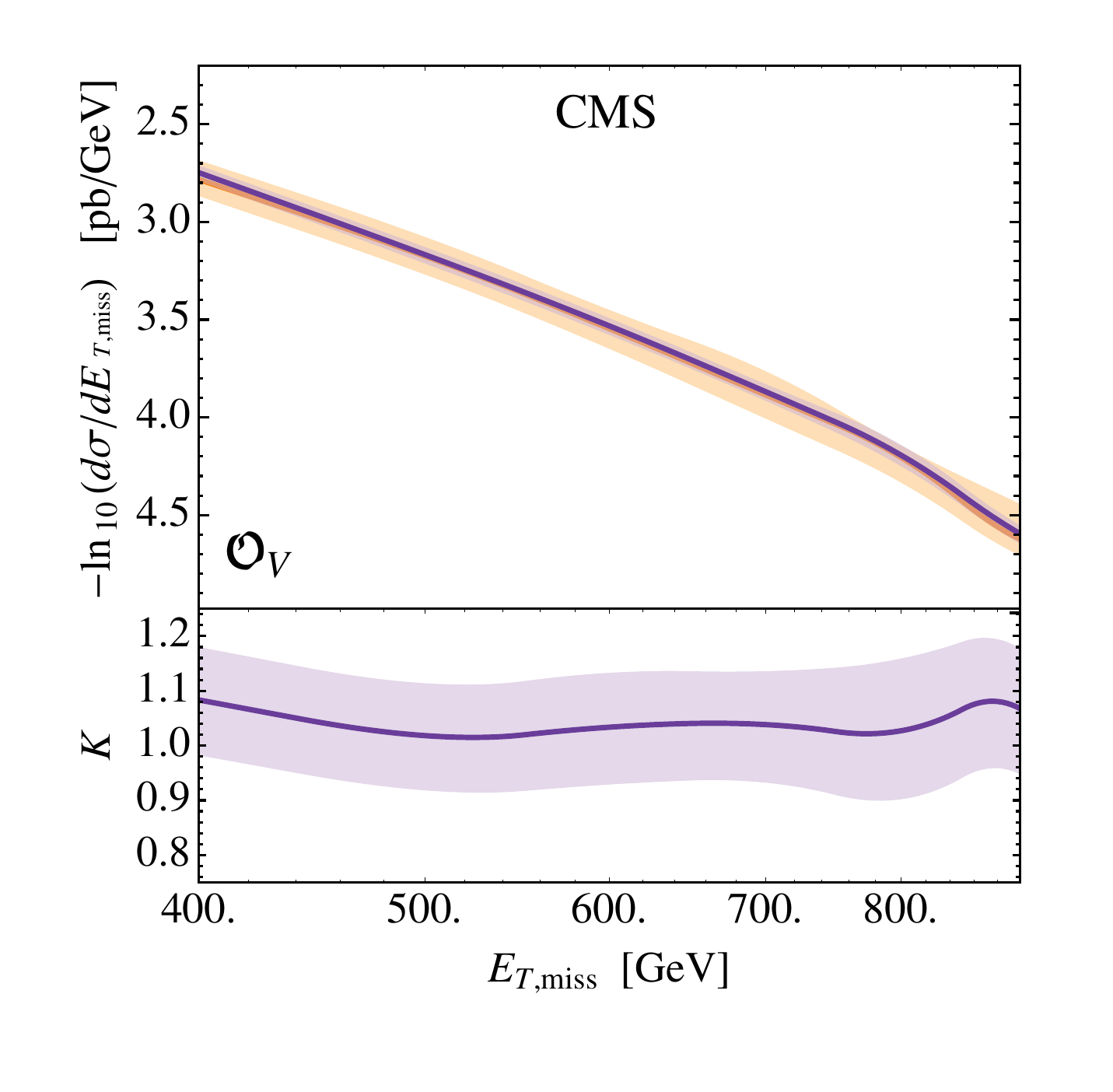}   \hspace{-4mm} 
\includegraphics[width=0.5\columnwidth]{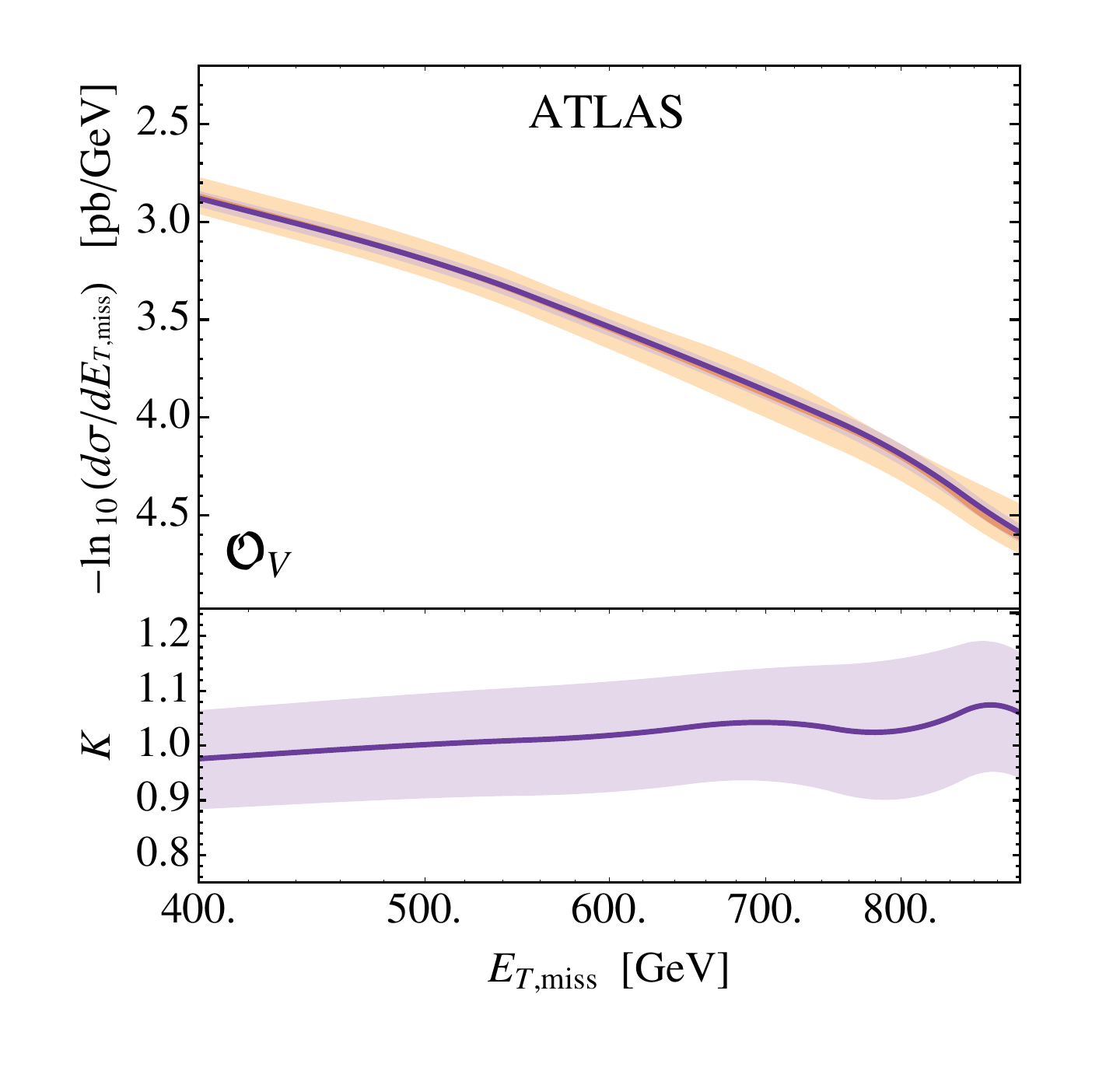}   
\end{center}
\vspace{-8mm}
\caption{\label{fig:ET} NLOPS (purple  curve and band) and LOPS (orange curve and band) predictions for the $E_{T,{\rm miss}}$ spectrum and the corresponding $K$ factor. The left (right) plot shows the  vector operator results obtained for $m_\chi = 10 \, {\rm GeV}$ employing the CMS (ATLAS) event selection criteria with a jet veto.}
\end{figure}

For completeness, we also study the effect of the NLO corrections on the $E_{T,{\rm miss}}$ distributions. In figure~\ref{fig:ET} we plot the NLOPS and LOPS predictions for the differential mono-jet cross section for a fixed DM mass of $10 \, {\rm GeV}$. Like in the case of the total cross sections, we observe that the scale ambiguities are reduced by a factor of 2 and that the differences between the NLOPS and LOPS results are small. The  corresponding $K$ factors are  almost flat in $E_{T,{\rm miss}}$ and amount to $1.1$ and $1.0$ at CMS and ATLAS with variations of roughly~$10\%$ around the central values. Even smaller differences arise in the case of the $p_{T,j_1}$ spectra.

\begin{figure}[t!]
\begin{center}
\includegraphics[width=0.5\columnwidth]{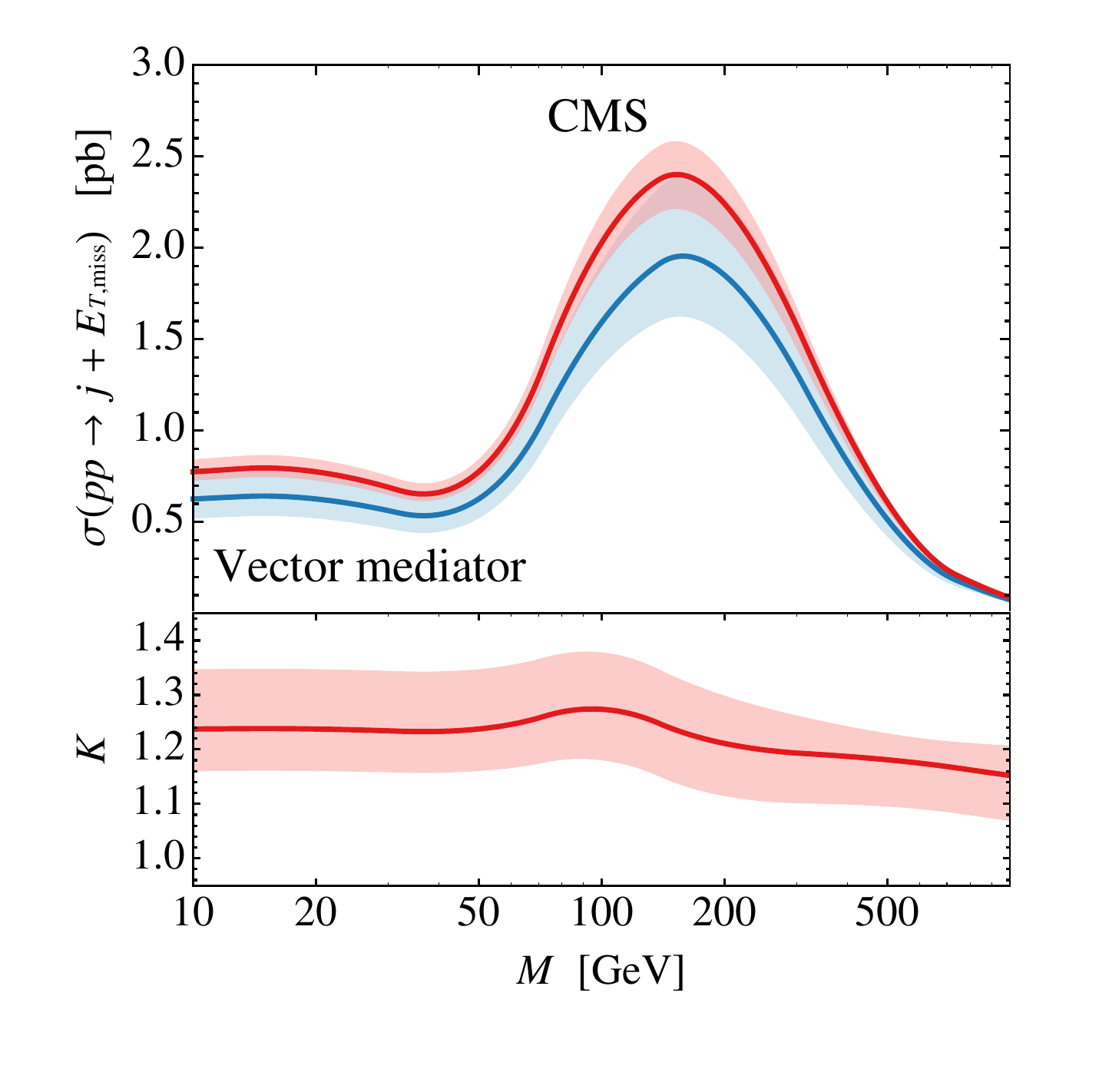}   \hspace{-4mm} 
\includegraphics[width=0.5\columnwidth]{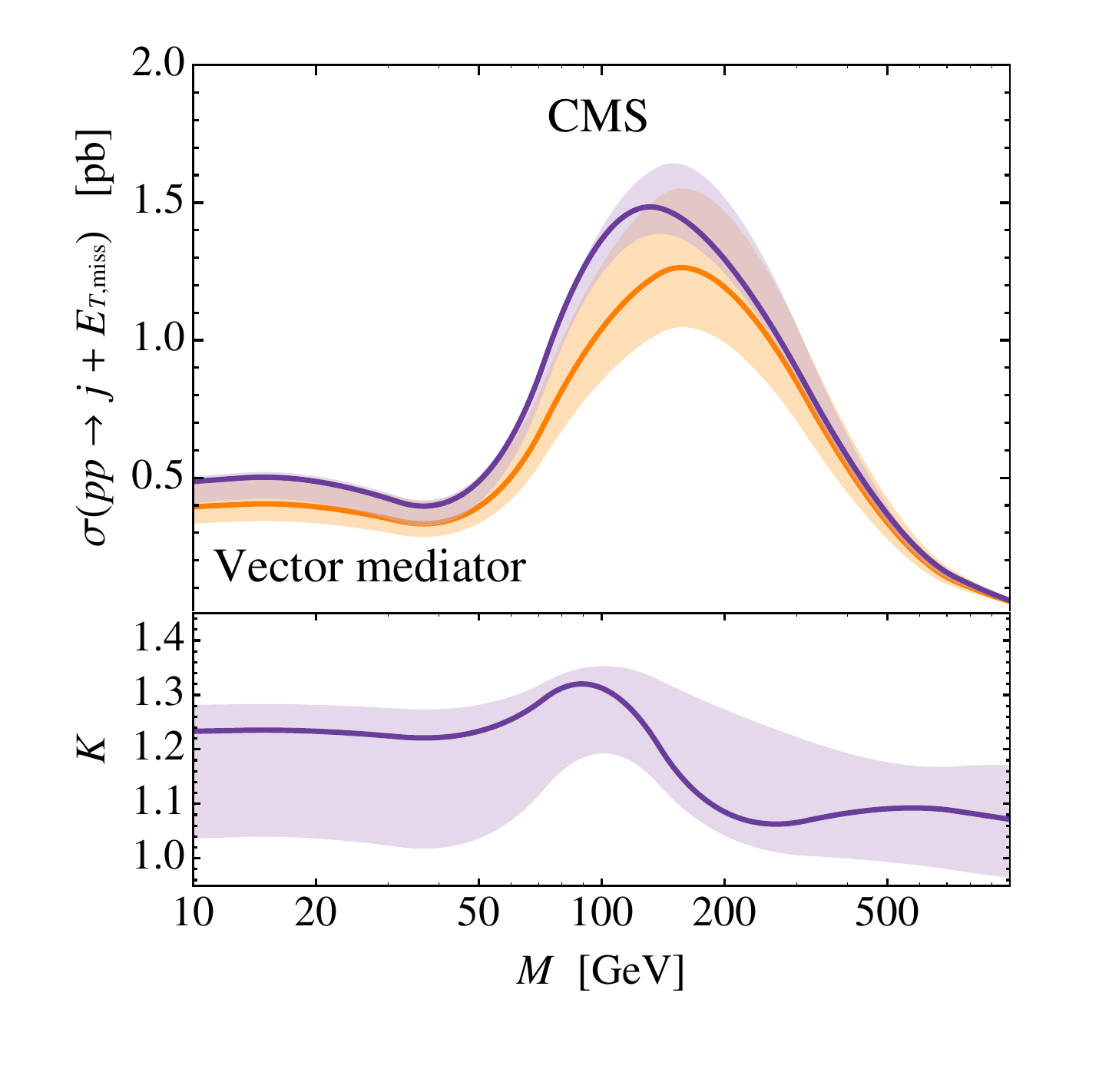}   
\end{center}
\vspace{-8mm}
\caption{\label{fig:resonance} Left panel: Fixed-order LO (blue) and NLO  (red) predictions for the jet + $E_{T, {\rm miss}}$ cross section as a function of the mediator mass $M$. Right panel: LOPS (orange) and NLOPS~(purple) cross sections with jet veto for different values of $M$. Both plots correspond to vector-like DM-quark interactions and assume CMS cuts, a DM mass of $m_\chi = 50 \, {\rm GeV}$ and a total decay width of the mediator of $\Gamma = M/3$. }
\end{figure}

The fact that  the $K$ factors are flat as a function of $E_{T,{\rm miss}}$ and $p_T$ implies that they can to first approximation also be used beyond the EFT, in particular in cases when the mediator of the interaction is resonantly produced. Nevertheless, to allow for a more detailed study of this case, our {\BOX} extension is fully capable of simulating on-shell mediators. As an illustration, we plot the mono-jet cross sections and the $K$ factors as a function of the mediator mass in figure~\ref{fig:resonance}. The shown results are based on the CMS event selection and assume $m_\chi = 50 \, {\rm GeV}$ and $\Gamma = M/3$. From the left panel we see that in the case of ${\cal O}_V$,  fixed-order NLO effects enhance the LO cross section by around $^{+30\%}_{+15\%}$ with the  larger (smaller) $K$ factors occurring at low (high) values of $M$. As illustrated in the right panel, a similar behaviour is observed  after including PS effects. While for $M \lesssim 2 m_\chi$ one finds values for $K$  of $1.2$, for $M \gtrsim 2 m_\chi$ the EFT result (\ref{eq:KVCMSATLAS}) is essentially recovered. The largest $K$ factors always occur at $M = 2 m_\chi$, which corresponds to threshold production of the DM pair. 

Unsurprisingly,  the vector ${\cal O}_V$ and axial-vector ${\cal O}_A$ operators show very similar behaviours for what concerns the importance of NLO and PS effects, scale dependencies and resulting $K$ factors. Visible differences occur only for large values $m_\chi$ of the DM mass, where the mono-jet cross section of the axial-vector operator is always slightly below the one of the vector operator. Given the similarity of the ${\cal O}_V$ and  ${\cal O}_A$ predictions we do not show plots for the case of the axial-vector operator. 

\subsection{Gluonic operator}
\label{sec:G}

\begin{figure}[t!]
\begin{center}
\includegraphics[width=0.5\columnwidth]{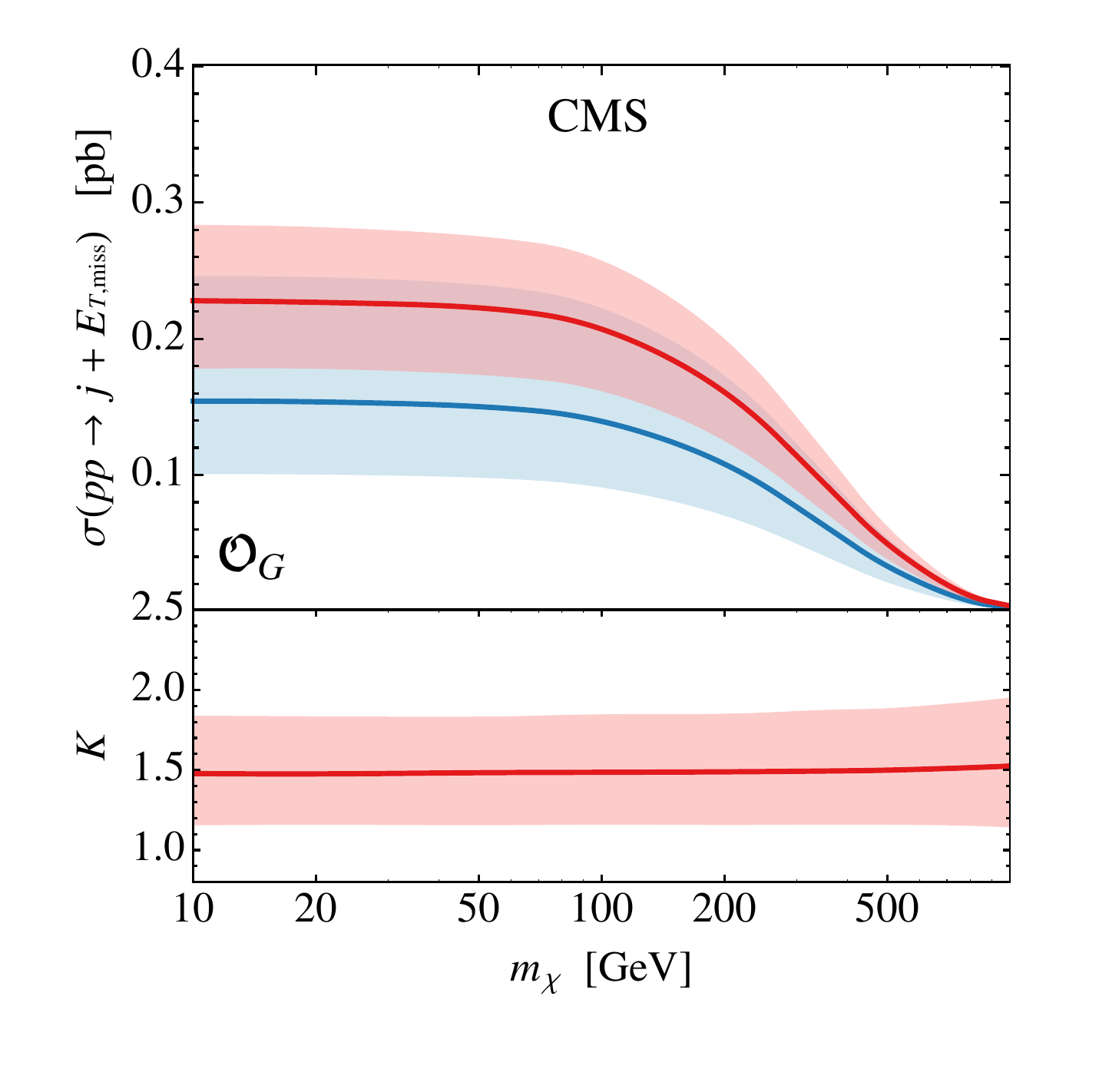}   \hspace{-4mm} 
\includegraphics[width=0.5\columnwidth]{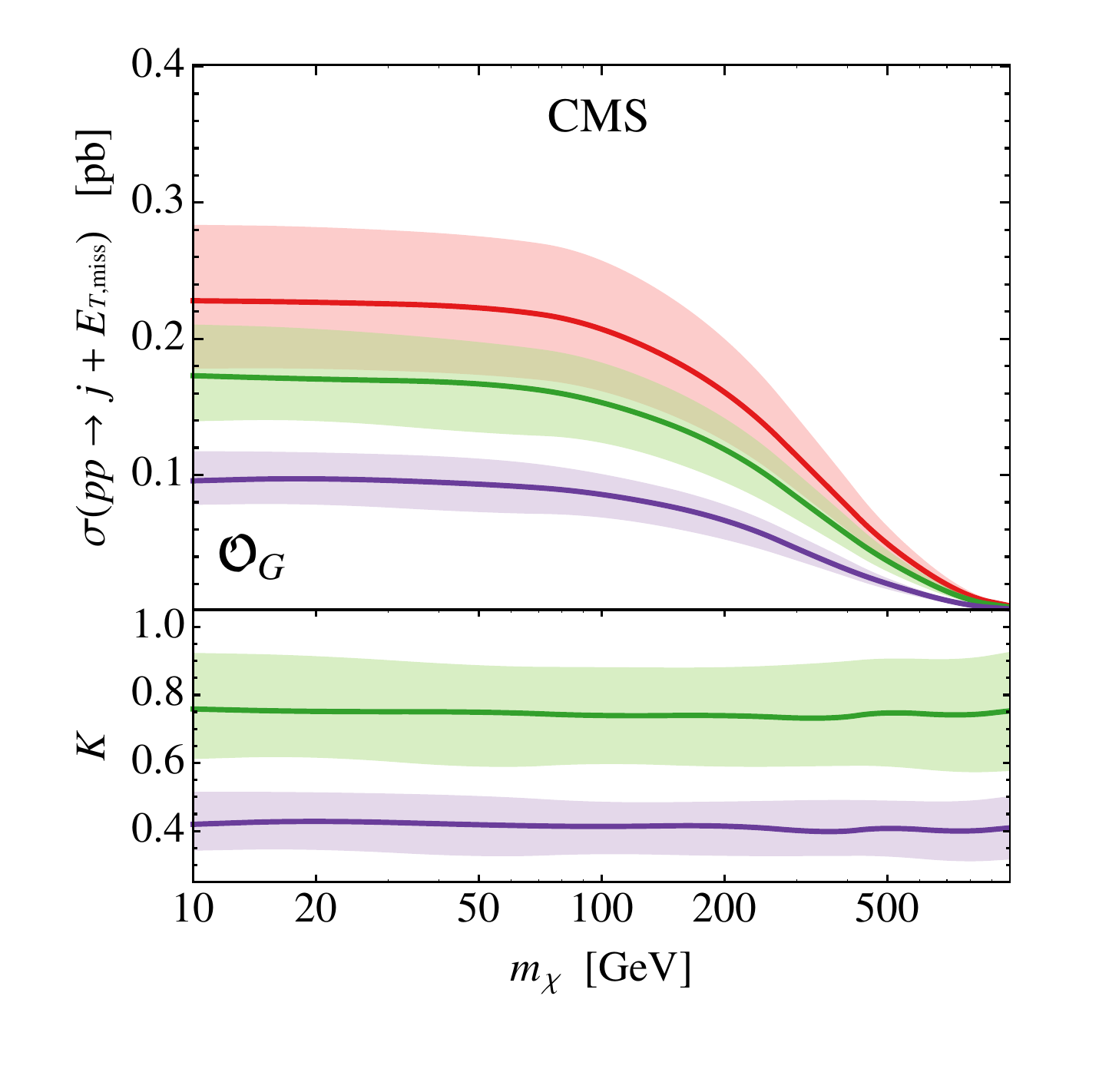}

\vspace{-6mm}

\includegraphics[width=0.5\columnwidth]{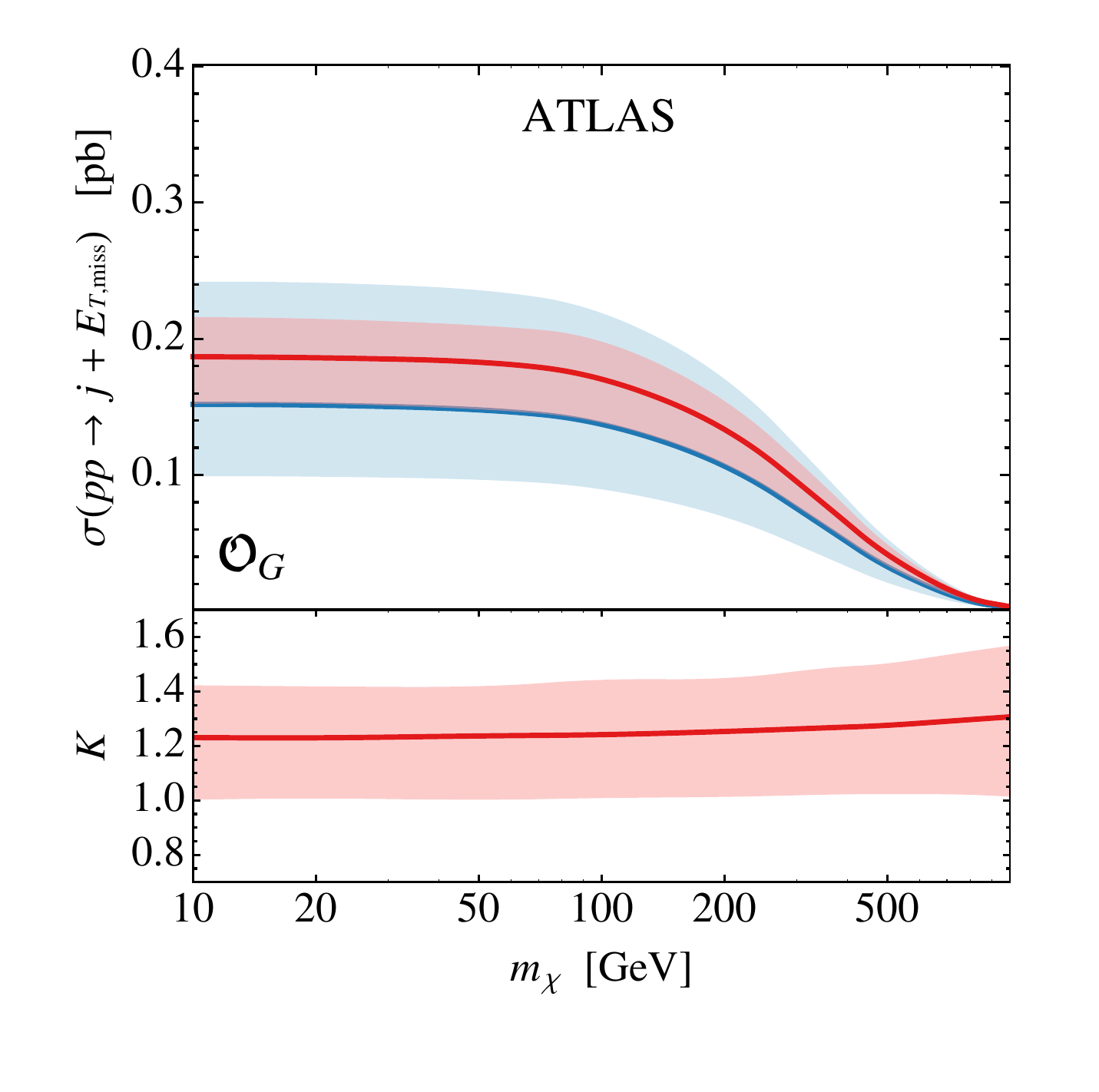}   \hspace{-4mm} 
\includegraphics[width=0.5\columnwidth]{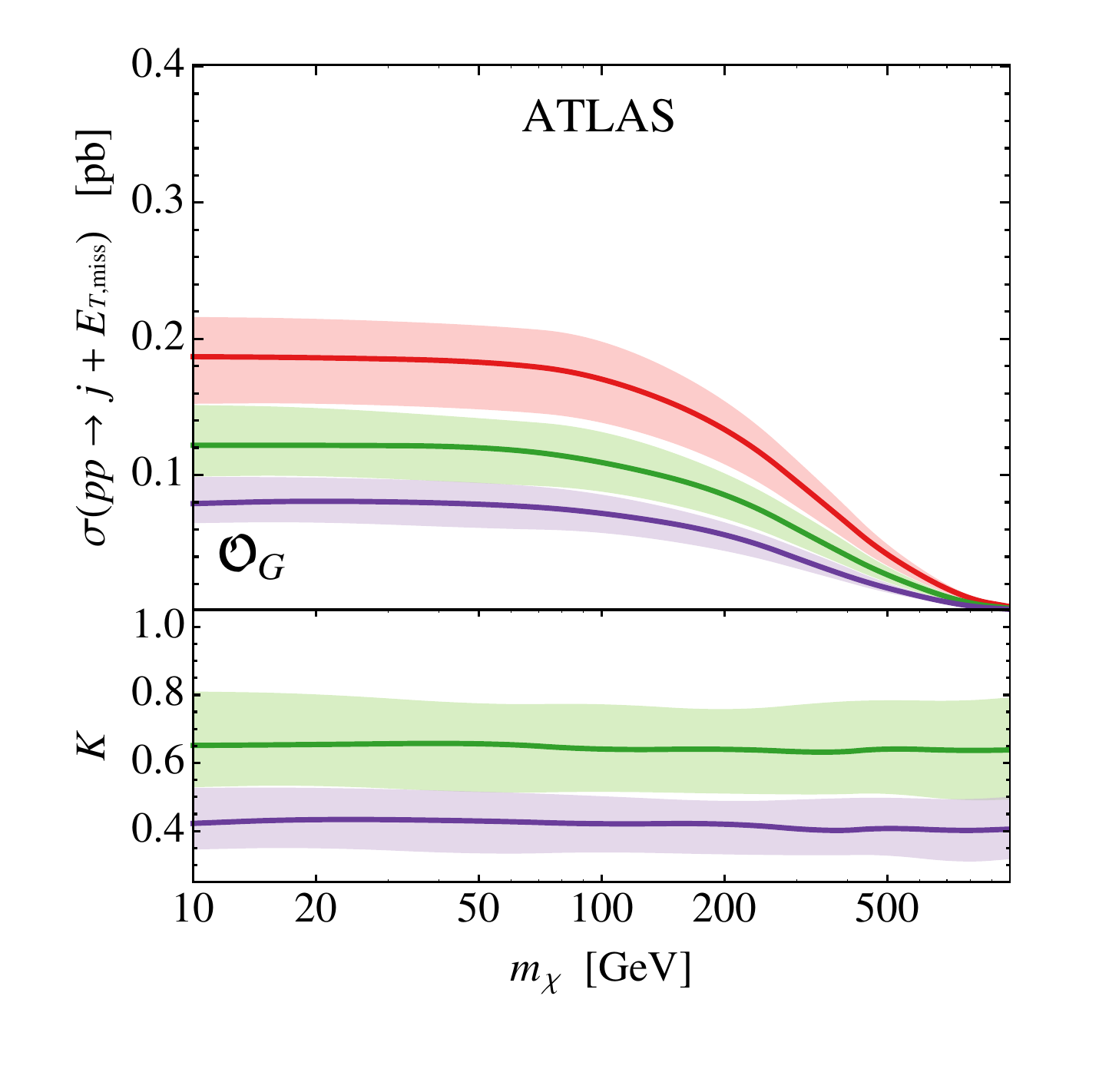}   
\end{center}
\vspace{-8mm}
\caption{\label{fig:CMSxsecG} Predictions for the gluonic operator ${\cal O}_G$ imposing the CMS (top) and ATLAS (bottom) event selection criteria. The colour coding agrees with the one in figure~\ref{fig:CMSxsec}.}
\end{figure}

Let us now consider the predictions for the gluonic operator ${\cal O}_G$ introduced in (\ref{eq:QGQGt}). In figure~\ref{fig:CMSxsecG} we compare the LO to the NLO results (left panels) and the NLO to the NLOPS predictions (right panels), imposing both CMS (top row) and ATLAS~(bottom row) cuts. As in the case of ${\cal O}_V$, we see that the fixed-order NLO corrections lead to larger cross sections for asymmetric than for symmetric cuts on $p_{T,j_1}$ and $E_{T,{\rm miss}}$. 

The resulting $K$ factors are approximately 1.5 and 1.3 and hence larger than in the case of the vector operator. The larger $K$ factors follow from the fact that for ${\cal O}_G$ mono-jet processes necessarily involve a gluon in the initial state and gluons radiate more than quarks. The different radiation pattern  also explains why the reduction of scale uncertainties is less pronounced for ${\cal O}_G$ than ${\cal O}_V$. The ratio between the fixed-order NLO and the NLOPS cross section with jet veto amounts to around 0.4 for both CMS and ATLAS.

\begin{figure}[t!]
\begin{center}
\includegraphics[width=0.5\columnwidth]{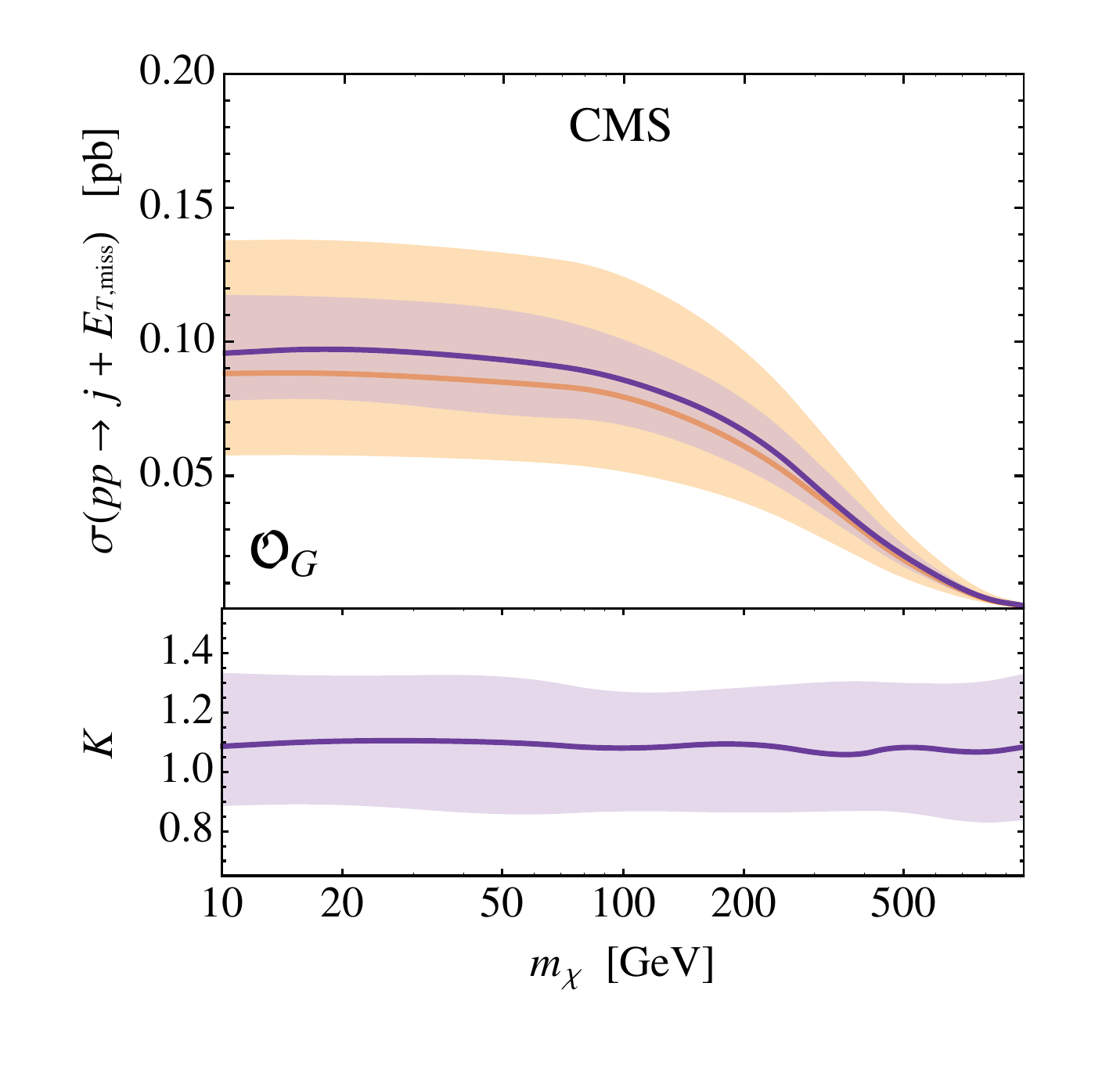}   \hspace{-4mm} 
\includegraphics[width=0.5\columnwidth]{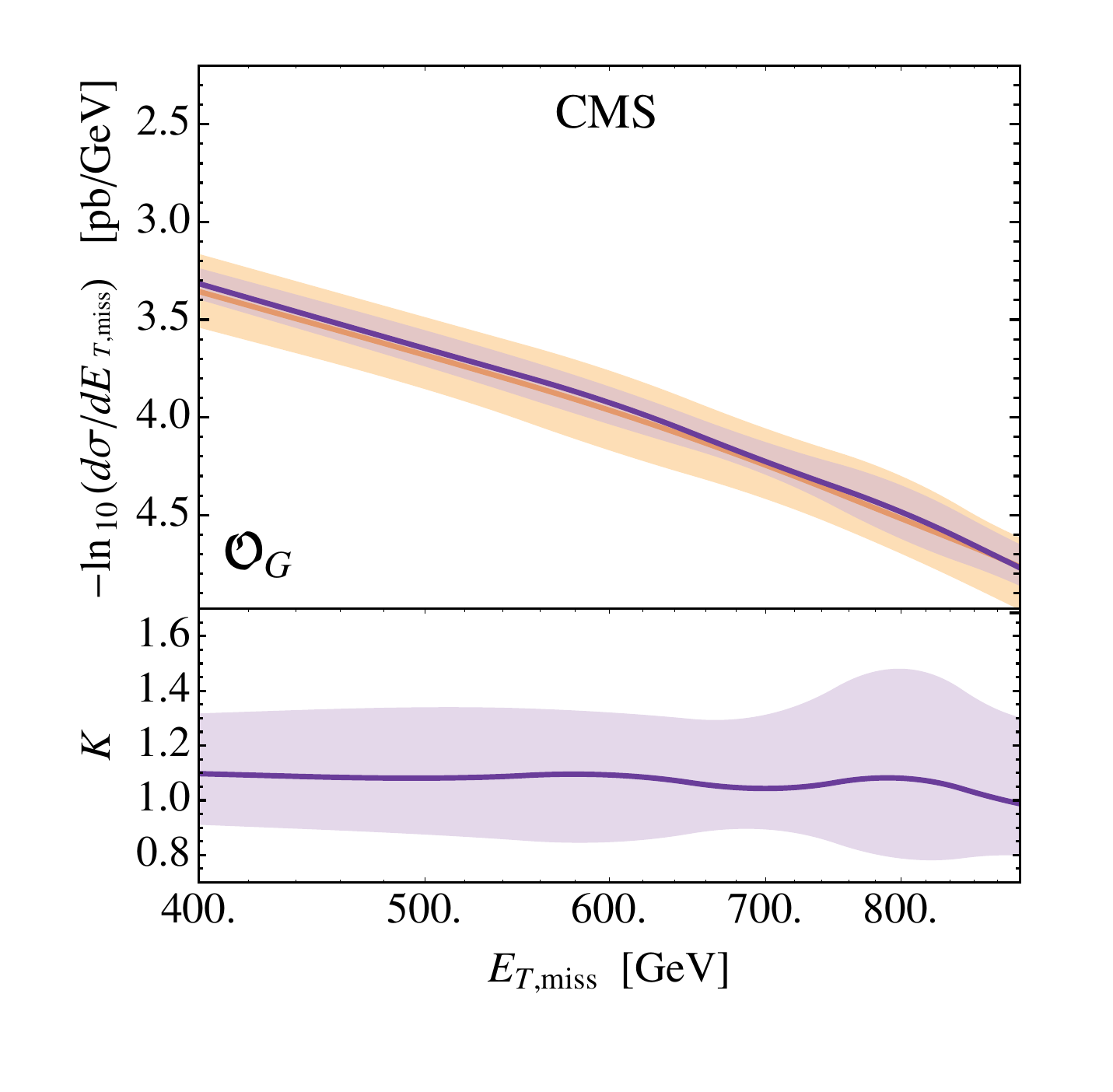}  

\vspace{-6mm}

\includegraphics[width=0.5\columnwidth]{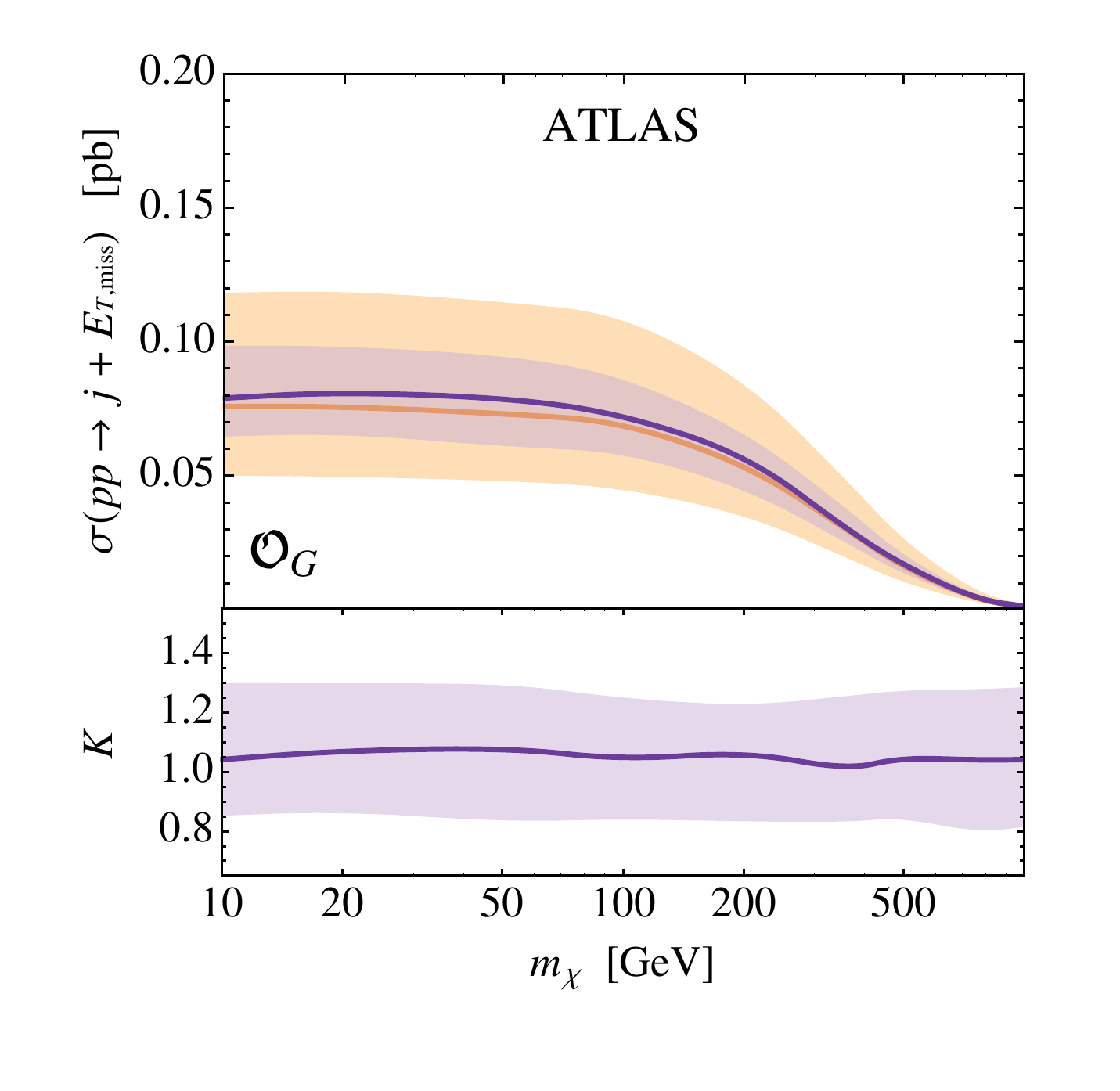}    \hspace{-4mm} 
\includegraphics[width=0.5\columnwidth]{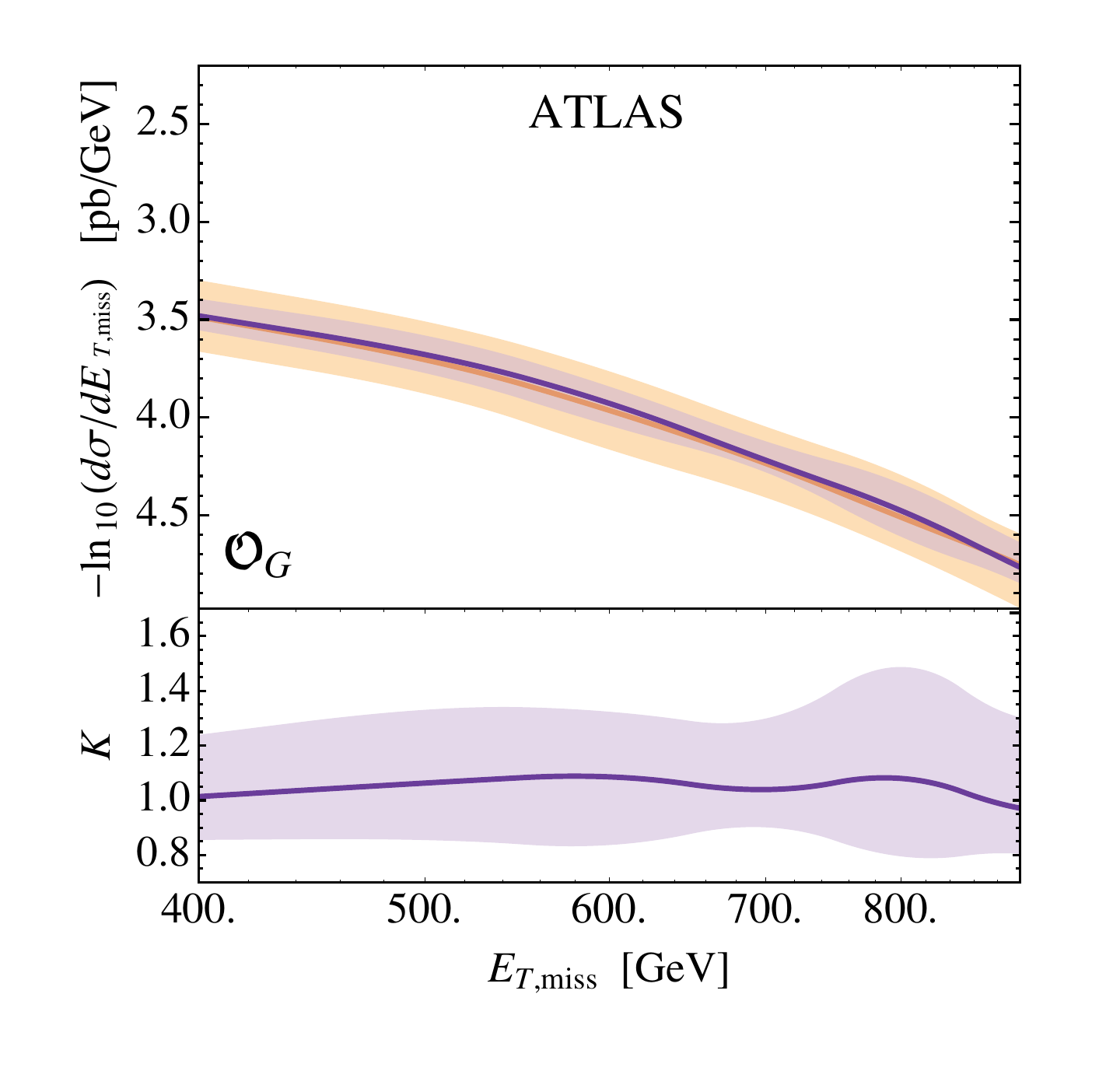}   
\end{center}
\vspace{-8mm}
\caption{\label{fig:realKG} Left: predictions for the gluonic operator ${\cal O}_G$ imposing CMS (top) and ATLAS~(bottom) cuts with $N_j \leq 2$. The colour coding is the same as in figure~\ref{fig:realK}. Right: predictions for the $E_{T,{\rm miss}}$ spectrum for  jet + $E_{T,{\rm miss}}$ production via ${\cal O}_G$.  The top~(bottom) panel shows the results corresponding to CMS (ATLAS) jet-veto cuts. The meaning of the coloured curves and bands is  analogue  to the one in figure~\ref{fig:ET}.}
\end{figure}

In figure~\ref{fig:realKG} we furthermore present a comparison of our NLOPS and LOPS predictions. We first observe that including higher-order QCD effects reduces the theoretical uncertainties by a factor of about 2. As for the vector operator the ratio of the NLOPS and the LOPS predictions are surprisingly close to 1.  We arrive at 
\begin{equation} \label{eq:KGCMSATLAS}
K_{\rm CMS}^G = 1.08^{+0.22}_{-0.22} \, , \qquad 
K_{\rm ATLAS}^G = 1.05^{+0.22}_{-0.21} \,.
\end{equation}
These numbers should be contrasted with the $K$ factors of 2 to 2.5 found in~\cite{Fox:2012ru} by comparing the fixed-order NLO and LO results. In order to better understand this  discrepancy we present in the middle pie chart of figure~\ref{fig:pies} the fractions of events with $N_j = 1, 2, 3$ and $N_j \geq 4$ jets. We see that for ${\cal O}_G$ merely 22\% of the signal in the CMS analysis is due to $1 \, {\rm jet} + E_{T,\rm miss}$ events, while in the remaining  $78\%$ cases the $E_{T,\rm miss}$ events involve multiple jets. This implies that soft radiation produced by the shower largely reduces the relative importance of the fixed-order NLO contributions, which in our case are found to be already smaller than in~\cite{Fox:2012ru}, due to our different scale choice.

\subsection{Scalar and pseudo-scalar operators}
\label{sec:PS}

\subsubsection{Impact of heavy-quark PDFs}

Since the scalar and pseudo-scalar operators (\ref{eq:QSQP}) involve quark-mass dependent couplings, their phenomenology is noticeably different from that of the vector and axial-vector operators discussed above. First of all,  because of the strong Yukawa suppression the resulting jet~+~$E_{T,{\rm miss}}$ cross sections are much smaller for ${\cal O}_S$ and ${\cal O}_P$ than those for ${\cal O}_V$ and ${\cal O}_A$.   As compensation we will use a smaller suppression scale of $\Lambda = 50 \, {\rm GeV}$ instead of  our reference value $\Lambda = 500 \, {\rm GeV}$ when analysing ${\cal O}_S$ and ${\cal O}_P$. Second, unlike the predictions for ${\cal O}_V$ and ${\cal O}_A$ that receive the by far largest contribution from light valence quarks in the initial state, in the scalar and pseudo-scalar cases DM pair production is dominated by processes with bottom and charm quark initial states~\cite{Haisch:2012kf,Fox:2012ru,Lin:2013sca}.\footnote{Note that, as discussed in section~\ref{sec:effective}, we do not include interactions between DM particles and top quarks in the scalar and pseudo-scalar operators.}

\begin{figure}[t!]
\begin{center}
\includegraphics[width=0.5\columnwidth]{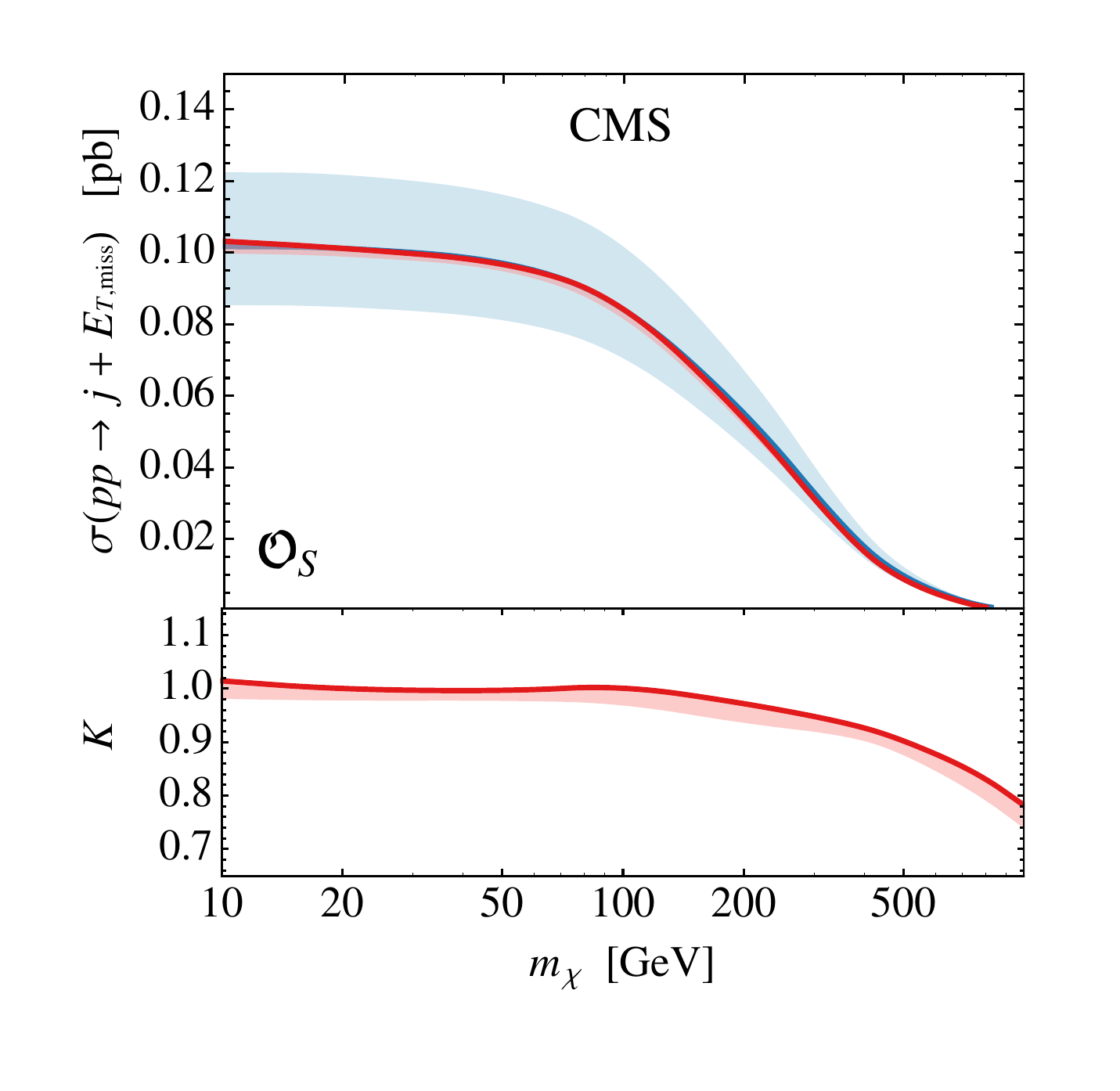}   \hspace{-4mm} 
\includegraphics[width=0.5\columnwidth]{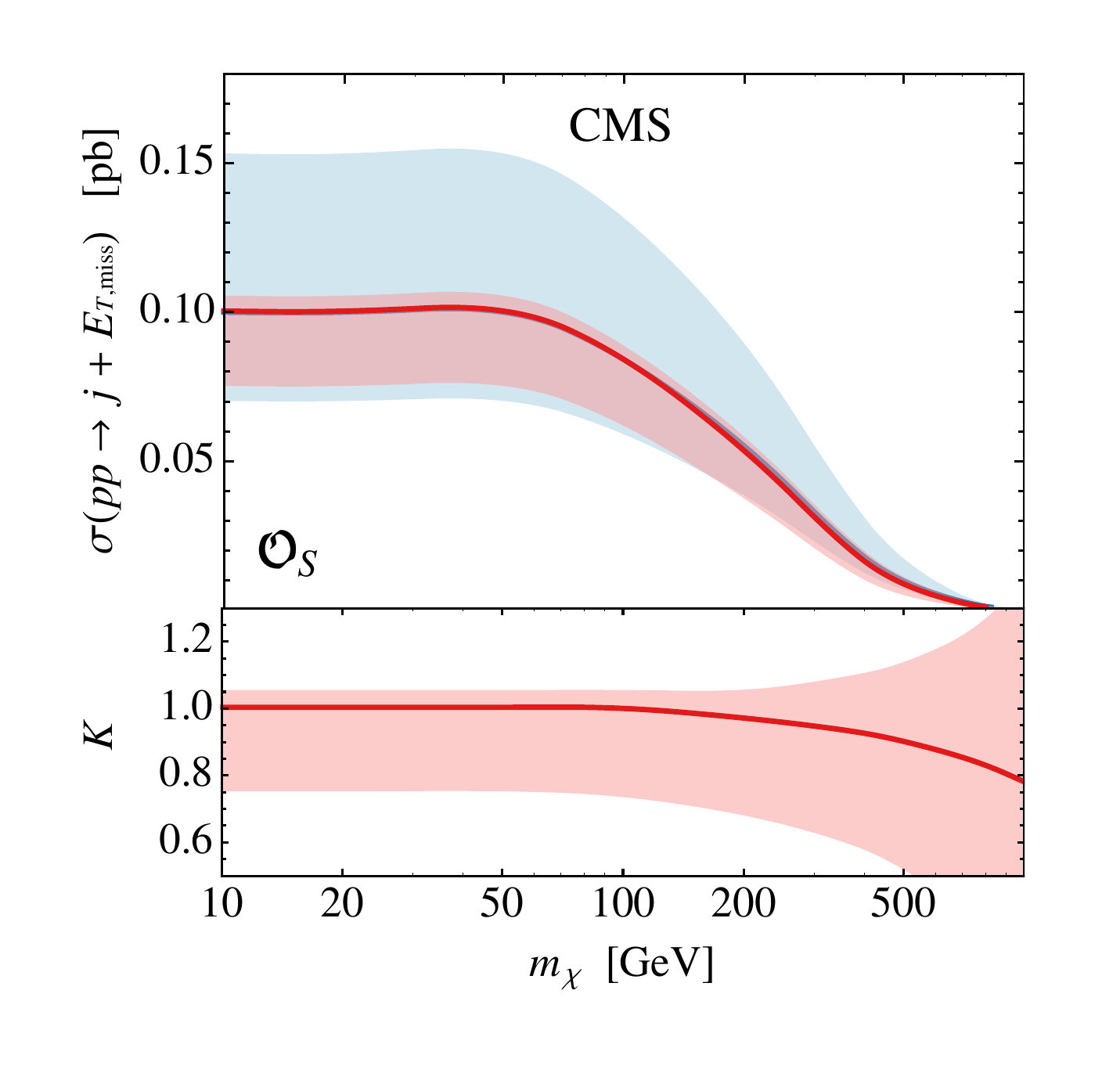}

\vspace{-6mm}

\includegraphics[width=0.5\columnwidth]{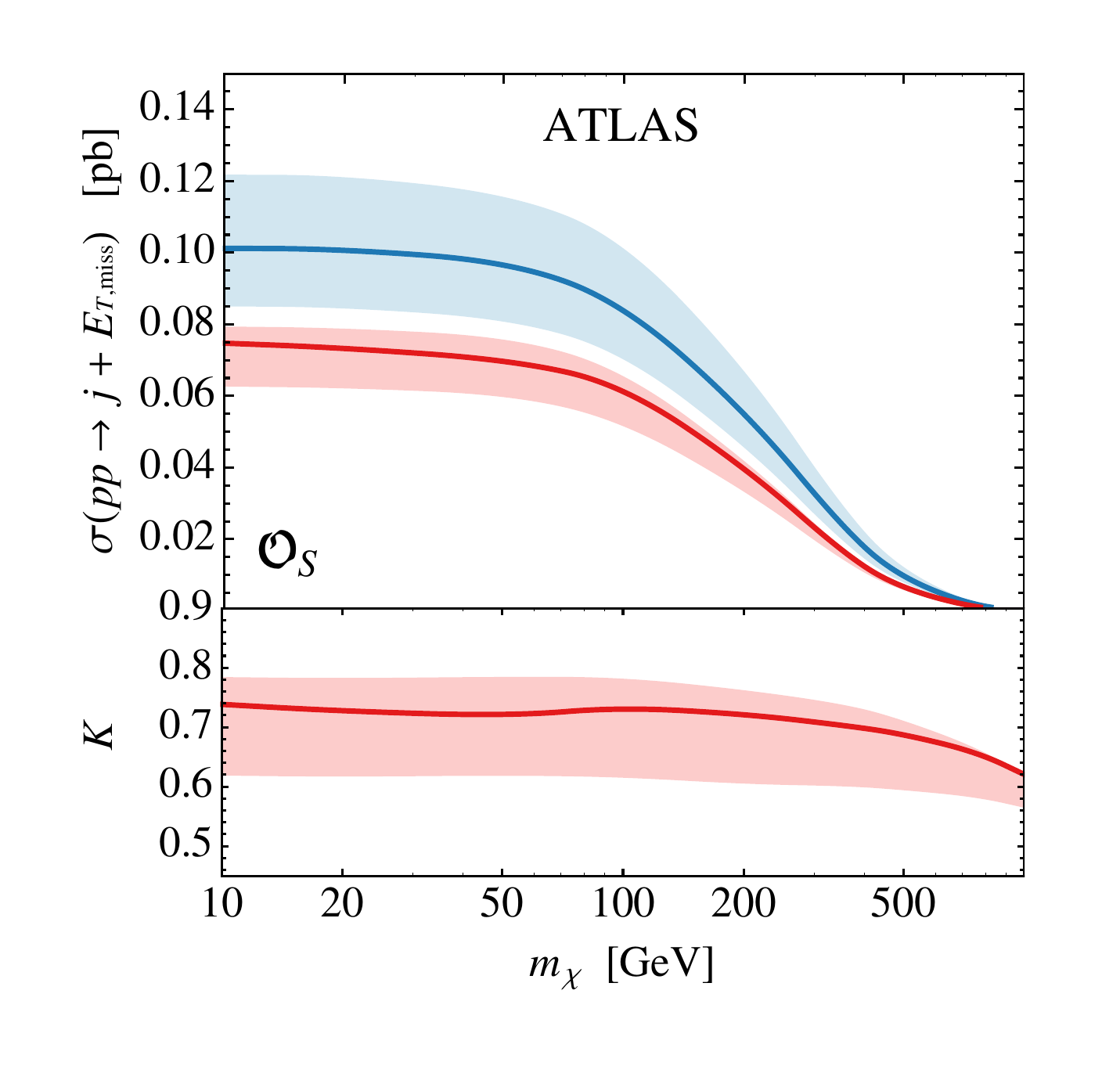}   \hspace{-4mm}
\includegraphics[width=0.5\columnwidth]{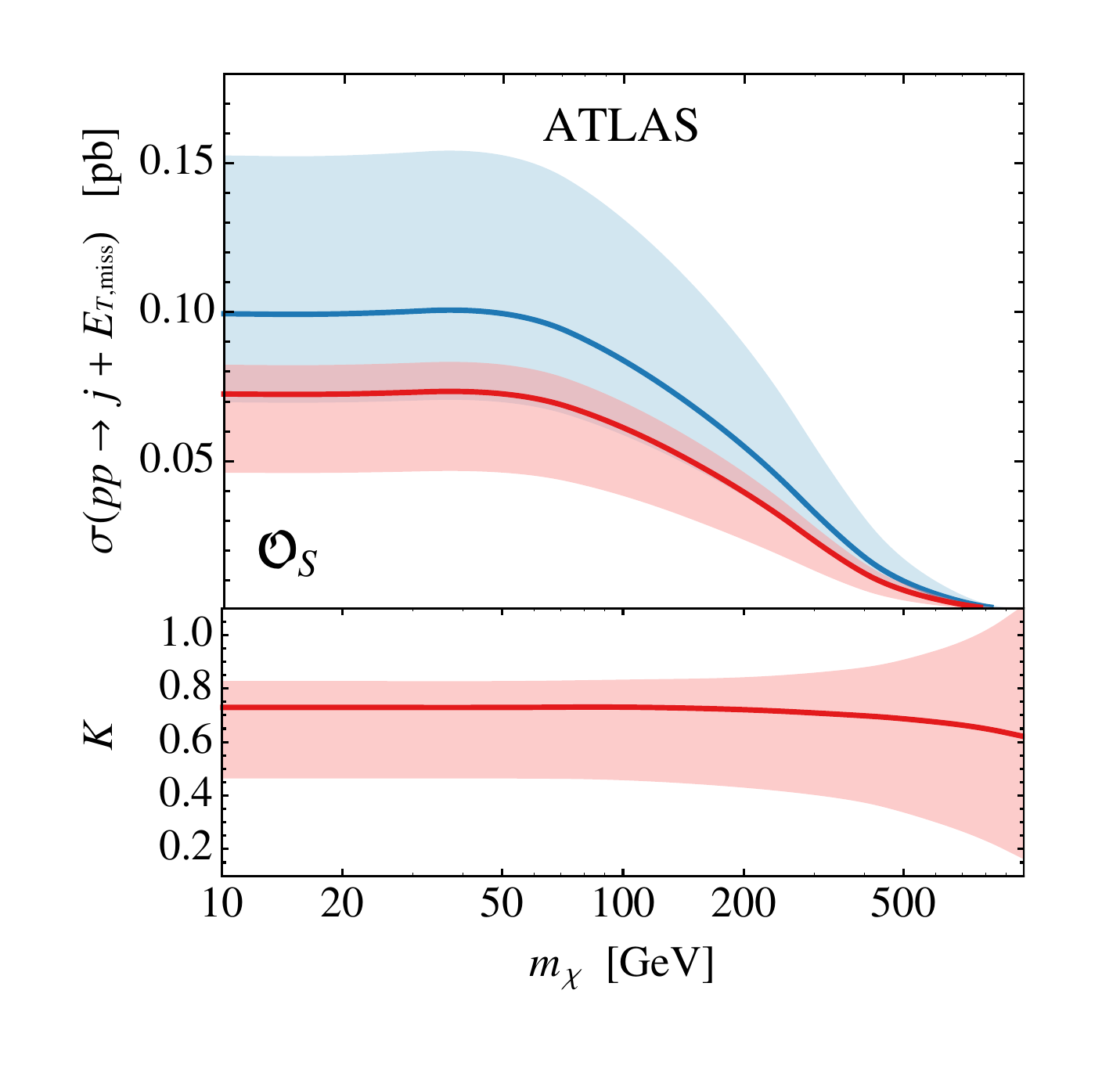}   
\end{center}

\vspace{-8mm}
\caption{\label{fig:CMSxsecS} Fixed-order predictions for the scalar operator ${\cal O}_S$ employing the CMS (top) and ATLAS~(bottom) cuts. The left panels show the impact of scale variations on the cross sections, while the right panels illustrate the theoretical uncertainties associated to the choice of heavy-quark PDFs. }
\end{figure} 

A proper description of heavy flavours in PDF analyses is an intricate issue that is however vital for precision measurements at the LHC.  Various prescriptions for the treatment of heavy quarks (so-called flavour schemes)  exist and are employed in modern PDF fits.  In order to assess the theoretical uncertainties associated to the choice of heavy-quark treatment, we calculate the  jet~+~$E_{T,{\rm miss}}$  cross sections arising for ${\cal O}_S$ and ${\cal O}_P$ using three different PDF families: the MSTW2008 sets \cite{Martin:2009iq}, which use the TR scheme \cite{Thorne:1997ga,Thorne:2006qt}, the CT10 heavy-quark distributions~\cite{Lai:2010vv},\footnote{In our LO calculation we employ  CT10 sets, while at NLO we use CT10nlo distributions.} which are based on a procedure called ACOT~\cite{Aivazis:1993pi,Collins:1998rz}, and the NNPDF~functions~\cite{Ball:2011mu,Ball:2012cx},\footnote{At LO we use NNPDF~2.1 distributions, while our NLO predictions are based on NNPDF~2.3 sets.} which apply the FONLL method~\cite{Cacciari:1998it,Forte:2010ta}. While all these implementations are general-mass variable flavour number schemes with explicit heavy-quark decoupling~\cite{Collins:1978wz} they differ in the precise way the perturbative expansion is organised.

We compare in figure~\ref{fig:CMSxsecS} the theoretical uncertainties resulting from scale dependencies using MSTW2008 PDFs~(left panels) with the variations of the predictions associated to the different choices of heavy-quark PDFs~(right panels) for central scales $\xi =1$. One sees that the scale ambiguities are always smaller than the errors related to the heavy-quark treatment.\footnote{The scale variations found at NLO in the  CMS case are again accidentally small.} It is also evident that the PDF uncertainties are not constant in $m_\chi$, but that they grow with increasing DM mass. This is related to the fact that the heavy-quark PDFs employed in our study differ most significantly in the region of large parton momentum fractions $x$.  While for large $x$  the CT10nlo PDFs for bottom quark  and charm quark  are larger than the corresponding CT10 distributions,  the opposite trend is observed in the MSTW2008 and NNPDF cases. The CT10 functions hence lead to the biggest $K$ factors, while we obtain the smallest values of $K$ employing  NNPDF distributions. 

Our findings should be contrasted with those of the article~\cite{Fox:2012ru} which quotes $K$ values between 2 and 2.5,  stating that the observed enhancements are due to DM pair production via gluon fusion, which in the scalar case only enters  at the NLO level. We  find that the large $K$ factors are dominantly an artefact of the specific choice of PDFs adopted in that work (i.e.~CTEQ6L1 at LO and CT10 at NLO) and are not due to the $gg \to b\bar b + E_{T, \rm miss}$ channel. The relative unimportance of the gluon fusion sub-channel can indirectly also  be inferred from the pie chart depicted on the right in figure~\ref{fig:pies}. The fraction of expected single-jet events at CMS for ${\cal O}_S$ is 30\%, which is closer to the case of the vector operator~(35\%) than to the gluon operator (22\%). The fact that the radiation pattern for the scalar operator resembles more closely vector interactions than gluonic interactions indicates that processes with quarks in the initial state dominate over gluon fusion.\footnote{This is even more obvious at ATLAS, where the fraction of events with  $N_j = 1$ amount to $44\%$ for ${\cal O}_S$, compared to 47\% for the vector operator and 31\% for the gluon operator.}

\subsubsection{Results including PS effects}

\begin{figure}[t!]
\begin{center}
\includegraphics[width=0.5\columnwidth]{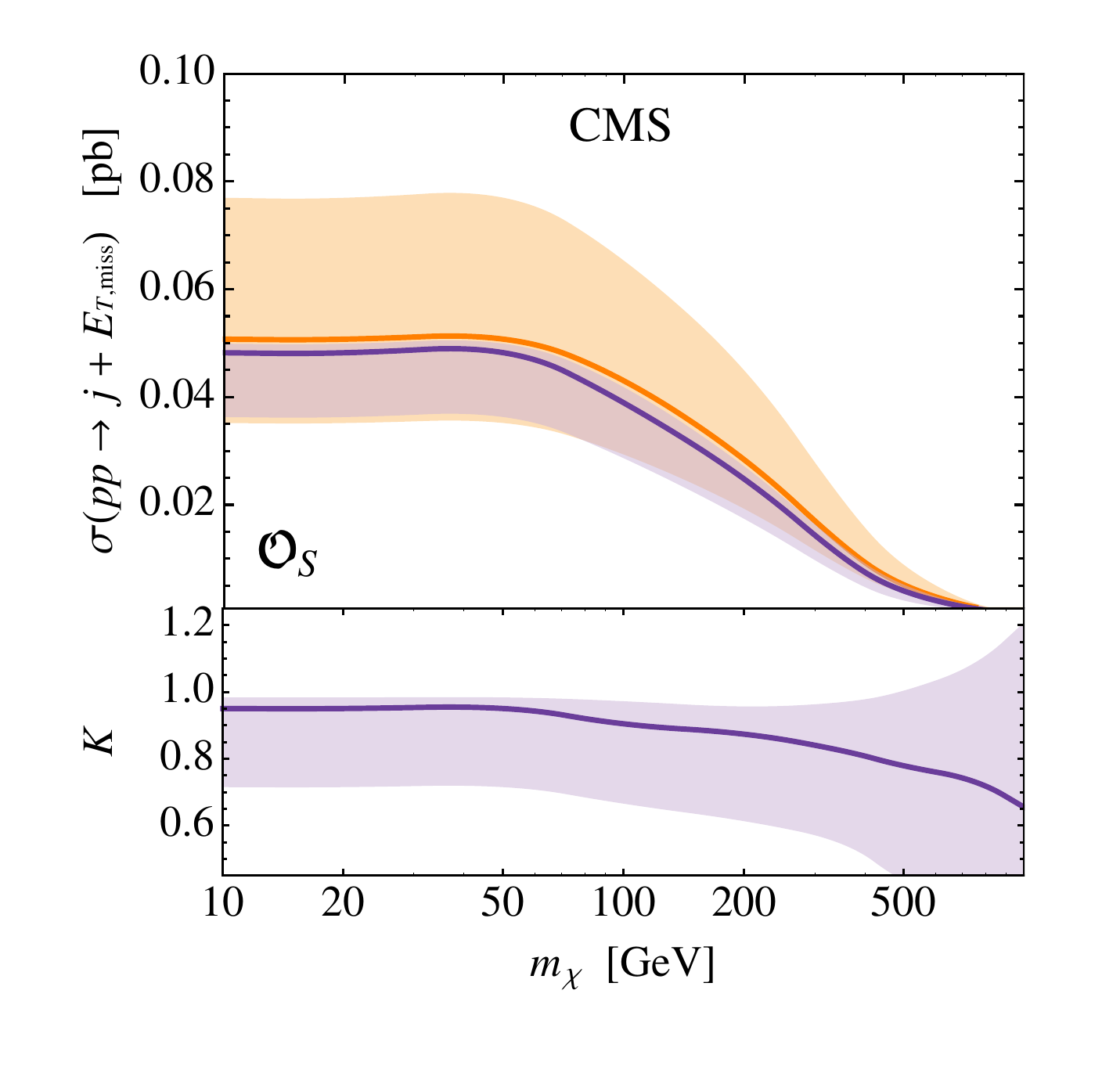} \hspace{-4mm} 
\includegraphics[width=0.5\columnwidth]{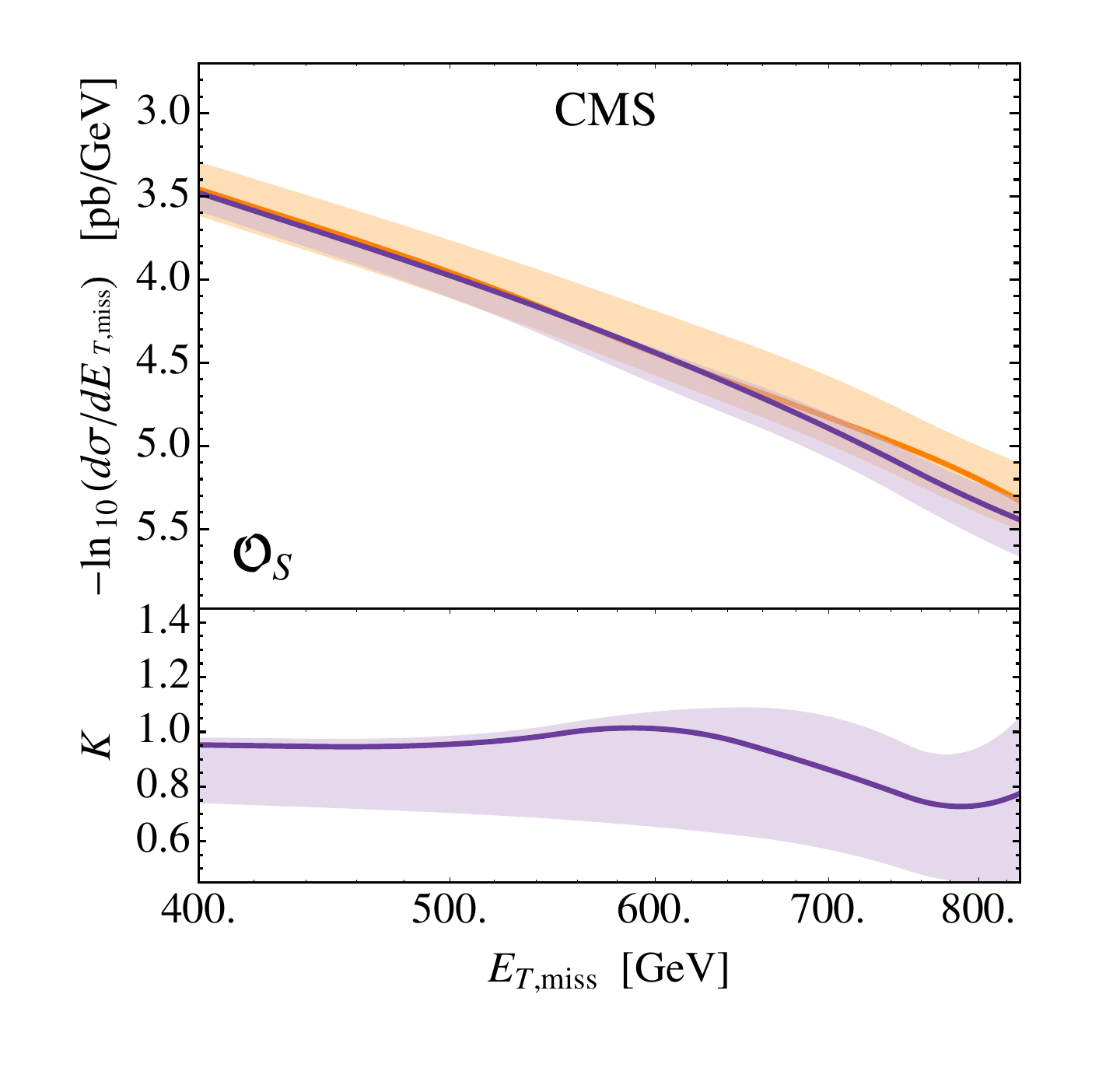} 

\vspace{-6mm}

\includegraphics[width=0.5\columnwidth]{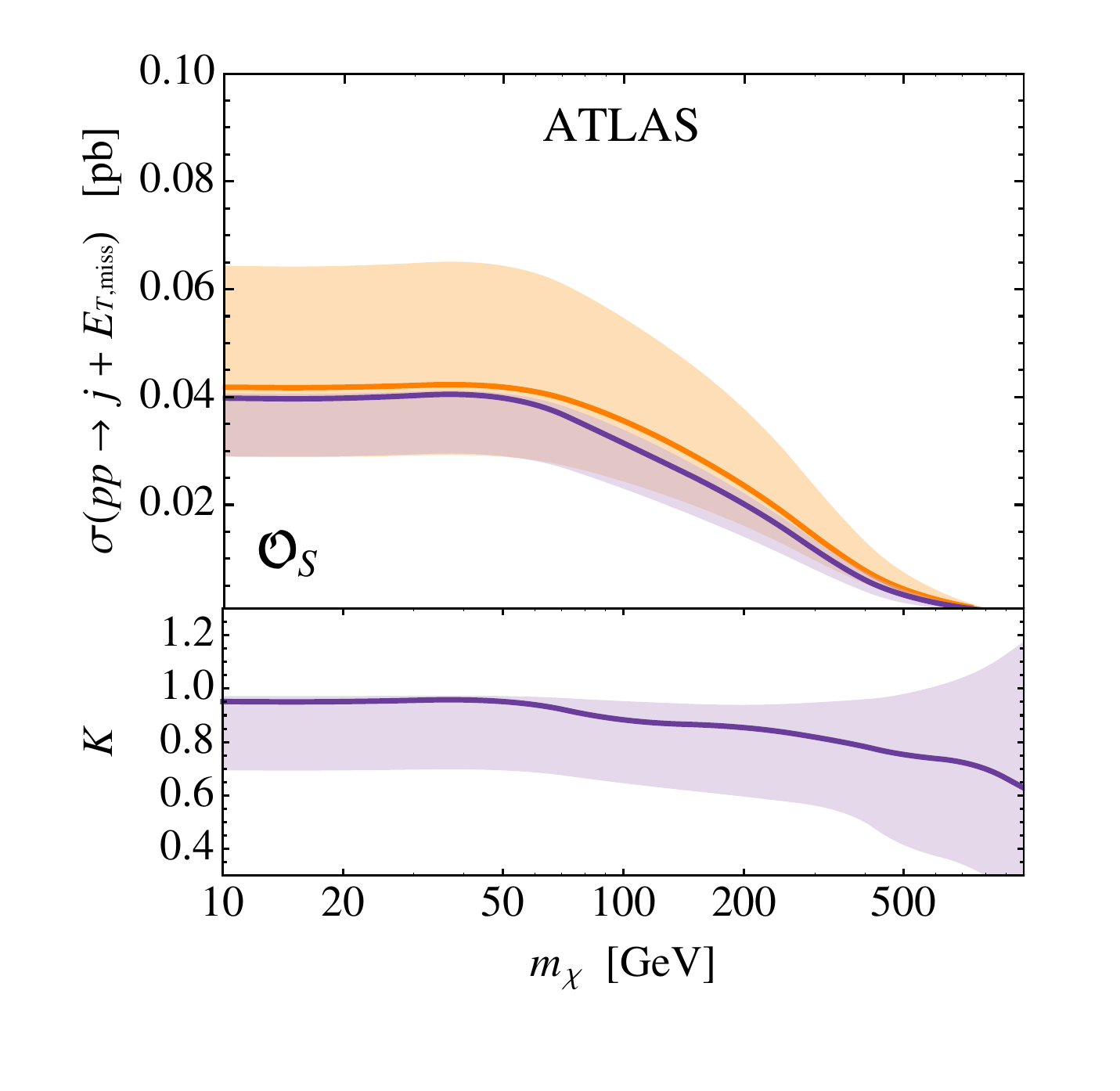}   \hspace{-4mm} 
\includegraphics[width=0.5\columnwidth]{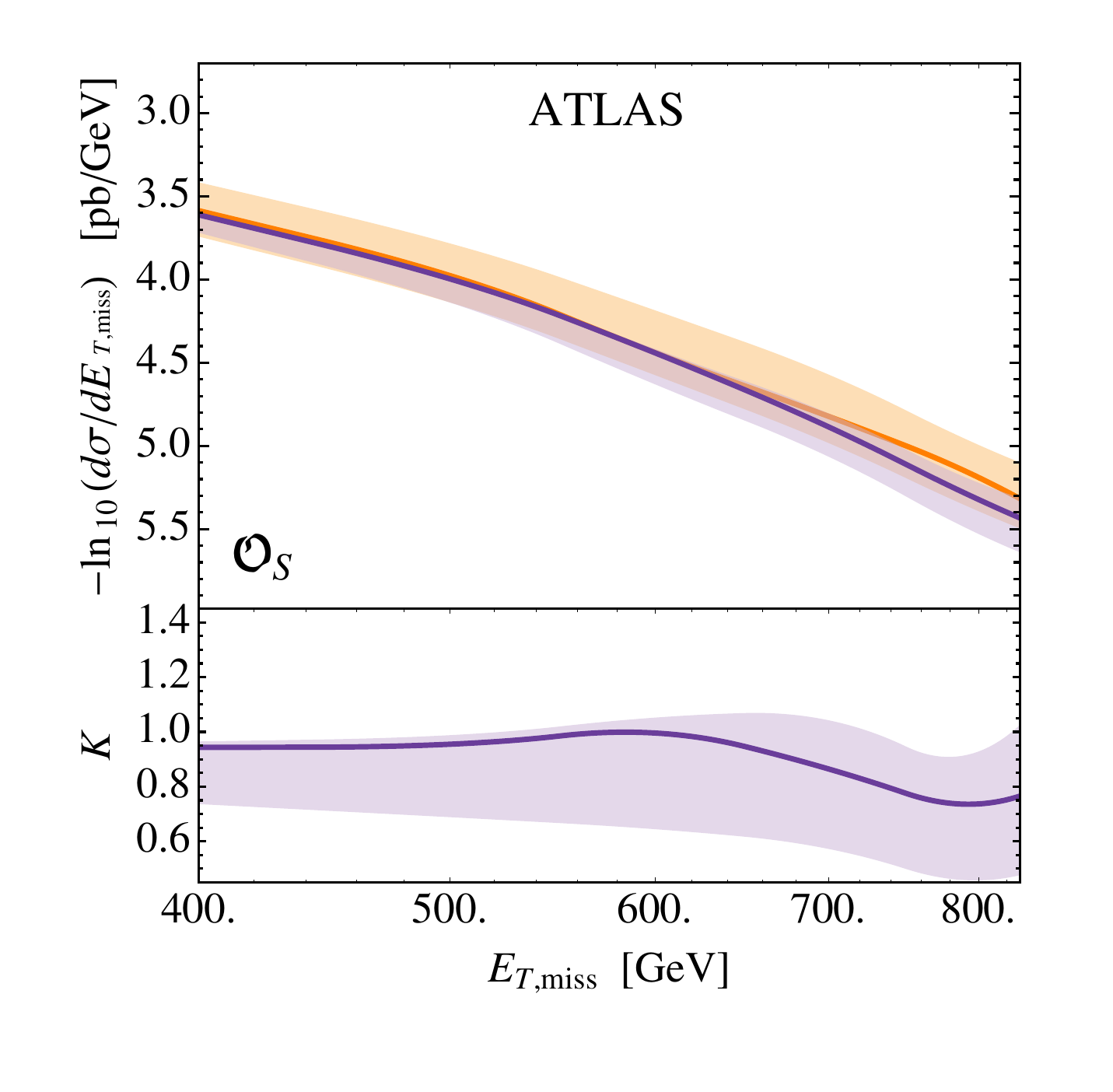}
\end{center}
\vspace{-8mm}
\caption{\label{fig:SNLOPS} Left: predictions for the scalar operator ${\cal O}_S$ assuming CMS (top) and ATLAS~(bottom) cuts with $N_j \leq 2$. The colour coding is the same as in figure~\ref{fig:realK}. Right: predictions for the $E_{T,{\rm miss}}$ spectrum for  jet + $E_{T,{\rm miss}}$ production induced by ${\cal O}_S$.  The top~(bottom) panel shows the results corresponding to CMS (ATLAS) jet-veto cuts. The colour coding  agrees with the one in figure~\ref{fig:ET}.}
\end{figure} 

We have seen above that both the scale ambiguities as well as the uncertainties related to the PDF choice can have a noticeable impact on the fixed-order predictions for ${\cal O}_S$. To obtain conservative error estimates of the cross sections we hence include both sources of uncertainties in our analysis of PS effects by determining the envelope of the various predictions. A comparison of our NLOPS and LOPS predictions for  ${\cal O}_S$ is given in figure~\ref{fig:SNLOPS}. As for the vector and gluonic operators we find that  including NLO effects reduces the theoretical uncertainties by a factor of approximately 2. Restricting ourselves to DM masses of $m_{\chi} \lesssim 100 \, {\rm GeV}$, we obtain the following $K$ factors
\begin{equation} \label{eq:KSCMSATLAS}
K_{\rm CMS}^S = 0.95^{+0.03}_{-0.24} \, , \qquad 
K_{\rm ATLAS}^S = 0.95^{+0.02}_{-0.25} \,, 
\end{equation}
where the asymmetry in the errors results from the PDF uncertainties. For heavier DM the uncertainty bands widen rapidly leading to $K$ factors of approximately $0.6 \pm 0.5$ for $m_\chi = 1 \, {\rm TeV}$. Similar observations apply in the case of the $E_{T,\rm miss}$ spectra which are shown on the right side in figure~\ref{fig:SNLOPS}. Within errors the $K$ factors that should be used to rescale the ${\cal O}_S$ results of existing experimental analyses are therefore compatible with 1. We however emphasised  that the choice of PDFs is important for ${\cal O}_S$.

The observations made above also hold in the case of the pseudo-scalar operator ${\cal O}_P$, which behaves very similar for what concerns the importance of higher-order QCD effects, scale dependencies and $K$ factors. In view of the similarity of the ${\cal O}_S$ and ${\cal O}_P$ predictions we do not show plots  for the latter case.

\section{Bounds on suppression scales}
\label{sec:bounds}

In the following we use the results from the preceding section to derive bounds on the suppression scale $\Lambda$ that enters the effective operators in (\ref{eq:QVQA}) to (\ref{eq:QGQGt}).  We emphasise that we do not attempt to set stronger bounds than those obtained in previous analyses. The aim of this section is simply to reproduce these results while making explicit the uncertainty of these bounds related to scale ambiguities.

In their most recent analysis with an integrated luminosity of $19.5\,\text{fb}^{-1}$ at $\sqrt{s} = 8\,\text{TeV}$, CMS observes a total of 8056 events compared to a SM expectation of $7875 \pm 341$~\cite{CMS:rwa}. At $95\%$ confidence level (CL) this result excludes new contributions to the mono-jet cross section in excess of 882 events. For the operator $\mathcal{O}_V$ this result corresponds to a bound $\Lambda > 800\,\text{GeV}$ for $m_\chi = 10\,\text{GeV}$.\footnote{Note that we have chosen to use the requirement $E_{T,{\rm miss}} > 350\,\text{GeV}$ in order to facilitate comparison with ATLAS. Slightly stronger bounds could be obtained for more stringent cuts~\cite{CMS:rwa}.}

\begin{figure}[!t]
\begin{center}
\includegraphics[width=0.57\columnwidth]{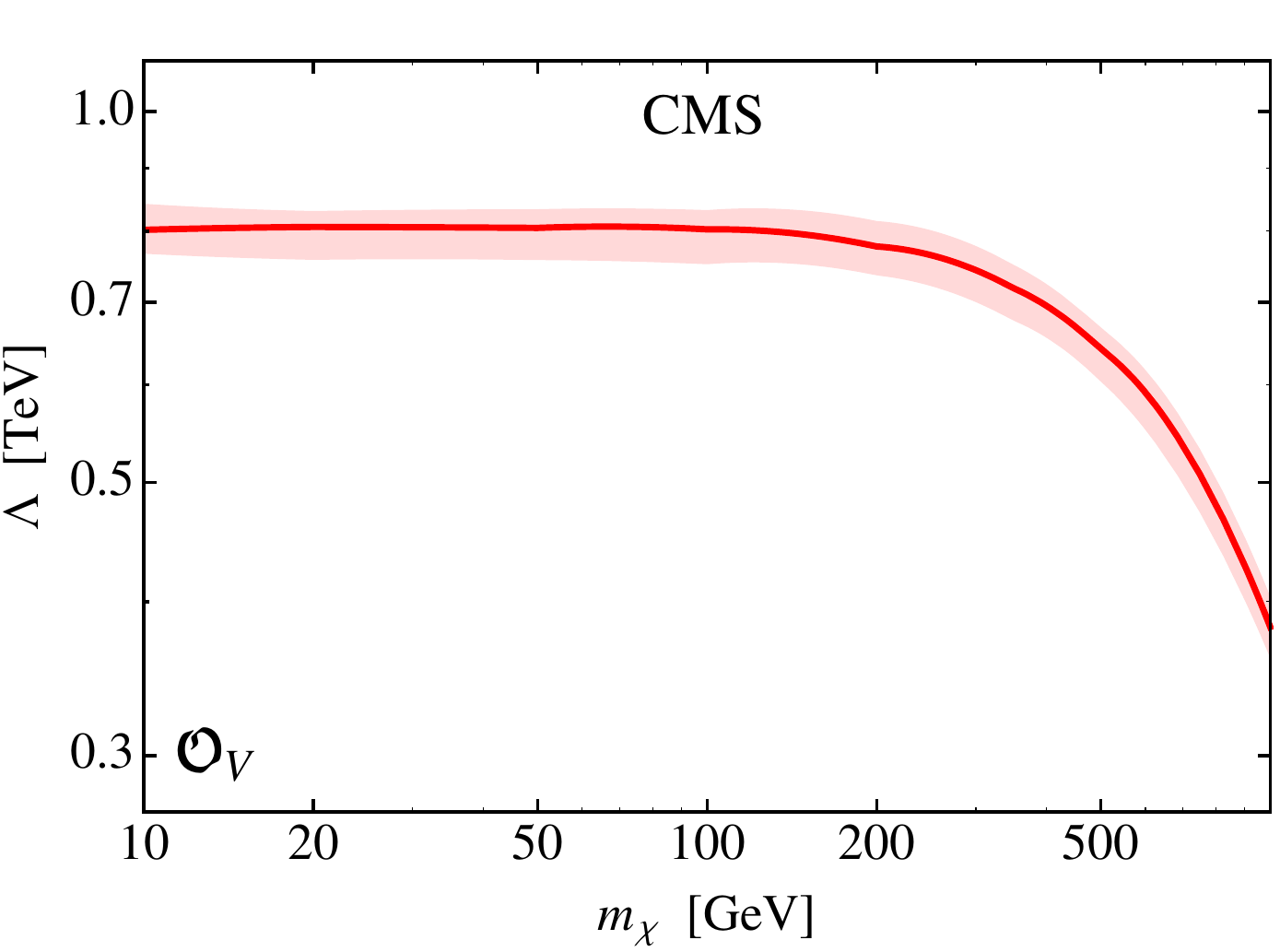}  
 
\includegraphics[width=0.57\columnwidth]{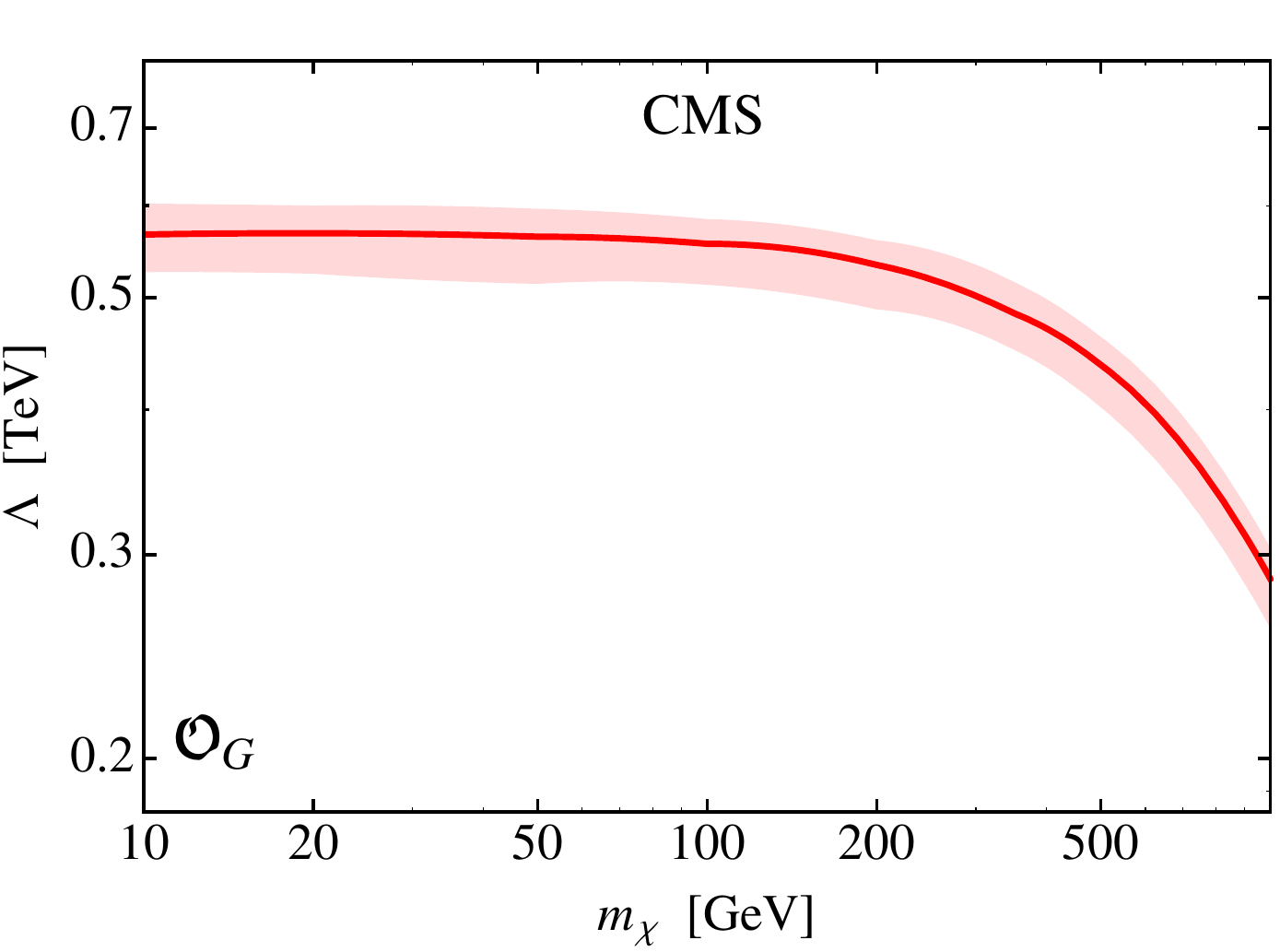}   

\includegraphics[width=0.57\columnwidth]{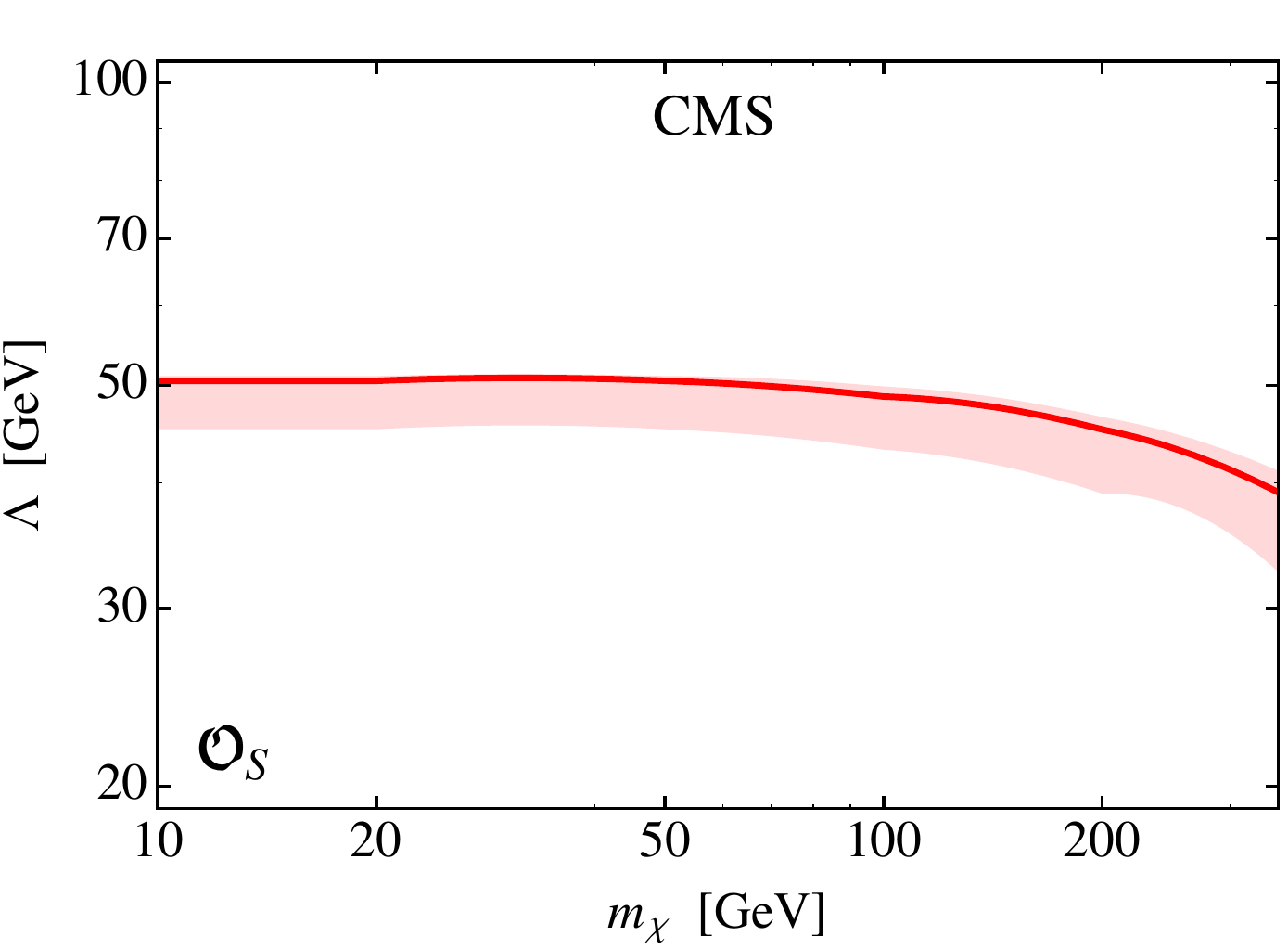}   
\end{center}
\vspace{-4mm}
\caption{\label{fig:bounds} Bounds on the suppression scale $\Lambda$ at 95\% CL as imposed by the latest CMS measurement. The top, middle and bottom panel shows the results for the vector operator ${\cal O}_V$, the gluonic operator ${\cal O}_G$ and the scalar operator ${\cal O}_S$, respectively.  The width of the bands indicates the theoretical uncertainties due to scale variations (as well as the choice of PDFs for ${\cal O}_S$).}
\end{figure}

According to figure~\ref{fig:realK} the mono-jet cross section at the LOPS level for the vector operator and $\Lambda = 500\,\text{GeV}$ is approximately $\big( 285^{+73}_{-54} \big) \,\text{fb}$ for small DM mass, which translates to  $5550^{+2780}_{-2070}$ expected events. Here we have multiplied the scale uncertainty by a factor of 1.96 to obtain a bound at $95\%$ CL. Since the number of events is proportional to~$\Lambda^{-4}$, we can therefore set a bound of $\Lambda > \big(792^{+85}_{-87}\big)\,\text{GeV}$ for small values of $m_\chi$. Our bound is in good agreement with the CMS result, even though we have not performed a detector simulation and used $R=0.4$ instead of the CMS jet radius $R=0.5$. We have however checked that the $R$ dependence of the mono-jet cross sections is weak. 
These findings indicate that the detector acceptance for mono-jet events should be very close to $100\%$.  This statement remains true for larger DM masses as we have explicitly verified. 

Having reproduced the CMS result at the LOPS level, we now turn to the NLOPS cross section. For this purpose, we make the assumption that the detector acceptance does not change significantly  if NLO corrections are included. Given the similarity of the LOPS and NLOPS cross sections, we believe that this is a very good approximation. For $m_\chi = 10\,\text{GeV}$, we then find $\Lambda = \big ( 802_{-35}^{+40} \big) \,\text{GeV}$ at $95\%$ CL. Had we simply used the averaged $K$ factors derived in section~\ref{sec:LONLO} instead, we would have obtained $\Lambda = \big ( 808_{-47}^{+35} \big)  \,\text{GeV}$, which agrees with the previous result within less than $1\%$. We conclude that the $K$ factors from above can be used to a very good approximation to estimate how strongly the limits on $\Lambda$ are affected by scale ambiguities. 

For the operators ${\cal O}_V$, ${\cal O}_G$ and ${\cal O}_S$ the resulting bands are shown in the  panels of figure~\ref{fig:bounds}. Note that our definition of the gluonic operator differs from~\cite{Goodman:2010ku} by a factor of~4. Consequently, our bounds on $\Lambda$ need to be rescaled by a factor of $4^{1/3} \approx 1.6$ to be compared to the results in~\cite{CMS:rwa}. As for the vector operator, we find good agreement also in the case of the gluonic and scalar operators between the different methods of calculating the limits on $\Lambda$. Numerically, we obtain for ${\cal O}_G$ a 95\%~CL bound of $\Lambda = \big ( 567_{-41}^{+36} \big) \,\text{GeV}$, while in the case of ${\cal O}_S$ we arrive at  $\Lambda = \big ( 50_{-4}^{+2} \big) \,\text{GeV}$. Both bounds hold for light DM with $m_\chi \lesssim 100 \, {\rm GeV}$. Notice that in the case of the scalar operator ${\cal O}_S$, we do not show the limits on $\Lambda$ for $m_\chi > 350 \, {\rm GeV}$, since the large theoretical errors render them not very meaningful. Moreover, given the weakness of the bound on $\Lambda$, there are significant concerns regarding the validity of the EFT for such large DM masses~\cite{Shoemaker:2011vi,Busoni:2013lha,Profumo:2013hqa, Buchmueller:2013dya}.

\section{Conclusions}
\label{sec:conclusions}

In this article, we have studied DM pair production in association with one or more jets, leading to large amounts of missing transverse energy. We have extended the {\BOX}  framework to include such processes, in order to determine the effects of NLO corrections as well as parton showering. While most results presented in this work are based on an EFT approach for simplicity, the code is fully capable of considering cases where the details of the interaction can be resolved and the $s$-channel mediator can be produced on-shell.

While previous works have found a significant enhancement of the (parton-level) cross section due to NLO corrections, we observe that these enhancements are significantly reduced once the effects of parton showering are included. The reason is that events with two jets at parton level are very likely to end up producing three or more jets after parton showering because of the large CM energy of the process. Such events are rejected once we impose the cuts currently used by ATLAS and CMS in their mono-jet searches, leading to a significant drop in the cross section. While this effect could be ameliorated by making mono-jet searches more inclusive (i.e.~by allowing additional jets), such a modification would increase experimental backgrounds dramatically. These findings underline the importance of including PS when studying NLO corrections, as can be done in the {\BOX} framework.

For the cuts used most recently by ATLAS and CMS, we find that the cross section at NLOPS is always very similar to the LOPS cross section, which has been used by the experimental collaborations to set bounds on the couplings of DM to quarks and gluons. Nevertheless, we find that including NLO corrections significantly reduces the scale uncertainties of the signal prediction and therefore makes these bounds much more reliable. We provide both the ratios of NLOPS to LOPS cross sections and the corresponding scale uncertainties in such a way that they can easily be used together with existing experimental analyses. In particular, we find that for an appropriate scale choice, these quantities have only a very weak dependence on the mass of the DM particles and the kinetic variables of the process. To a very good approximation the presented ratios can therefore be used as overall factors to rescale the LOPS predictions. We demonstrate the usefulness of this approach by calculating the bounds on the suppression scale $\Lambda$ for three different effective operators. We emphasise that rescaling the LOPS cross section by a fixed-order $K$ factor would in general give wrong results.

Furthermore, we find that for symmetric cuts on $p_{T,j_1}$ and $E_{T,\text{miss}}$~---~as currently employed by the ATLAS collaboration~---~the impact of the PS on the results is more pronounced than in the asymmetric case. It thus seems favourable to have asymmetric cuts allowing for smaller values of $p_{T,j_1}$. On the other hand, if the cut on $p_{T,j_1}$ is too weak, the total cross section is dominated by events generated by the PS that are only LO accurate. In order to minimise theoretical uncertainties, the cut on $p_{T,j_1}$ should therefore be only slightly weaker than the cut on $E_{T,\text{miss}}$. Making this statement more precise will require a close collaborative effort between theorists and experimentalists.

With an increase in CM energy the LHC will soon make significant progress in constraining or discovering the interactions of DM particles with quarks and gluons. However, an increase in energy also implies larger QCD radiation effects  and a greater need for accurate theoretical predictions of the expected signal. The  {\BOX} extension presented in this paper will be a helpful tool for optimising experimental cuts, interpreting the data and extracting reliable bounds.

\acknowledgments{We are grateful to Steven Schramm and Giulia~Zanderighi for their valuable comments on the manuscript. We thank Patrick Fox, Tongyan Lin, Mario Martinez, Bjoern Penning,  Tim~Tait, James Unwin, Lian-Tao Wang and Steven Worm for useful discussions. UH and FK thank the KITP in Santa Barbara for hospitality and acknowledge partial support from the National Science Foundation under Grant No.~NSF PHY11-25915. FK would also like to thank the KICP at the University of Chicago for hospitality during the workshop ``Dark Matter at the LHC''. ER thanks the ESI in Vienna for hospitality while part of this work was carried out.  FK is supported by the Studienstiftung des Deutschen Volkes and a Leathersellers' Company Scholarship at St~Catherine's College, Oxford.} \\

\begin{appendix}

\section{POWHEG BOX implementation}
\label{sec:powheg}

In this appendix we give some technical details on how the mono-jet cross sections have been implemented into the \BOX. The processes in which we are interested, when considered at Born level, are of the type $ \bar\chi\chi\, +\, 1$~jet, with $ \bar\chi\chi$ being a fermionic pair produced by an $s$-channel exchange of a  colourless  spin-0 or spin-1 resonance $X$. Although in the EFT approach the intermediate resonance has been integrated out generating the higher-dimensional operators discussed in section~\ref{sec:effective}, as far as strong interactions are concerned, these processes are all of the type $X(\to \bar\chi\chi) + 1$~jet. The structure of these cross sections is hence very similar to that of $Z/H+1$~jet production in the SM, which allows us to build on the pre-existing \BOX{} infrastructure for these processes. To implement the DM production mechanism associated with ${\cal O}_{V,A}$ we extended the $Z+1$~jet code, while we modified $H+1$~jet production to deal with ${\cal O}_G$. The code associated to the (pseudo)-scalar operators ${\cal O}_{S,P}$ was written from scratch, since within the SM, $H + 1$~jet production involving a light-quark Yukawa coupling  is, given the  tininess of this channel, typically not considered.

Despite the aforementioned similarities, one should notice that for the processes at hand one needs to keep the exact DM mass dependence in the amplitudes. This does not pose a  conceptual problem since the production and decay part of the amplitudes factorise in the case of $s$-channel exchange. However, since the fermionic current associated to $\chi$ has to be considered massive, we had to extend the structure of  $Z+1$~jet production as originally implemented in the \BOX. From the technical point of view, instead of using massive spinors, we used the prescriptions outlined in~\cite{Rodrigo:2005eu,Badger:2010mg}, similarly to what has been done in~\cite{Fox:2012ru}, with the difference that in our code all the tree-level amplitudes entering Born and real corrections are constructed using Hagiwara-Zeppenfeld helicity amplitudes~\cite{Hagiwara:1985yu,Hagiwara:1988pp}.\footnote{As a by-product, this new implementation also allows for a NLOPS simulation of $Z+1$~jet production with the exact mass dependence retained in the $Z$-boson decay sub-amplitude. This can be relevant for precision studies of $Z\to \bar bb$ and, to a minor extent, for $Z\to \tau^+ \tau^-$.}

The virtual corrections for the processes involving the operators ${\cal O}_{V,A}$ and ${\cal O}_{G}$  resemble those of $Z/H+ 1$~jet production in the SM. In consequence, for the gluonic case we just needed to rescale the amplitudes as implemented in the \BOX{} $H+1$~jet code.  In the case of the vector and axial-vector operators,  we instead had to contract the one-loop helicity amplitudes with the DM current, including all possible helicity combinations. We used the results presented in Appendix~B of~\cite{Fox:2012ru} as a starting point and carried out all necessary algebraic manipulations with the help of \SatM{}~\cite{Maitre:2007jq}. The final results were written in terms of spinor chains, whose numerical evaluation is performed using the MCFM machinery. The same strategy was used to obtain the results in the case of  the scalar and pseudo-scalar operators ${\cal O}_{S,P}$. 

To validate our implementations we have performed various cross-checks. In the case~$m_\chi = 0$, we have checked that our $X(\to \bar\chi\chi) + 1$~jet amplitudes agree point-by-point  with the  \BOX{}  amplitudes for  $Z/H+ 1$~jet production, while for $m_\chi \neq 0$ we have compared our results with  the MCFM implementations  made available recently by the authors of \cite{Fox:2012ru}. A final successful cross-check was also carried out by comparing differential NLO distributions obtained with MCFM and our \BOX{} implementation.

\section{MC simulation of mono-jet signal}
\label{sec:manual}

In this appendix we describe step by step how to perform an MC
simulation of $p p \to \bar \chi \chi + j$ using our \BOX{}
implementation.  We restrict ourselves to the case of spin-1 mediators
and assume that the reader is familiar with the common features of the
\BOX{} package. These are explained in the user manual located in the
directory {\tt POWHEG-BOX/Docs} of the \BOX{} SVN repository.

\subsection*{Event generation}

The generation of events starts by building the relevant executable:\\[2mm]
{\tt \$ cd POWHEG-BOX/DMV} \\
{\tt \$ make pwhg\_main}\\

\noindent Hard events are then generated by \\[2mm]
{\tt \$ cd testrun-lhc}\\
{\tt \$ ../pwhg\_main}\\

\noindent After approximately 30 minutes of running time, the file {\tt pwgevents.lhe} will contain 50000 events for $p p \to \bar \chi \chi + j$  arising from an insertion of ${\cal O}_V$. In order to shower these events with PYTHIA 6, one has to execute\\[2mm]
{\tt \$ make main-PYTHIA-lhef }\\ 
{\tt \$ cd testrun-lhc} \\
{\tt \$ ../main-PYTHIA-lhef} \\

\noindent in the directory {\tt POWHEG-BOX/DMV}. Showering the
generated 50000 Les Houches events with PYTHIA takes just over 5
minutes. Events can also be showered using
HERWIG~\cite{Corcella:2000bw} (the corresponding file that one has to
build is called {\tt main-HERWIG-lhef}), PYTHIA
8~\cite{Sjostrand:2007gs} or Herwig++ \cite{Bahr:2008pv}.

\subsection*{Process specific input}
\label{sec:input}

The input parameters in the file {\tt
  POWHEG-BOX/DMV/testrun-lhc/powheg.input} that are specific to the $p
p \to \bar \chi \chi + j$ process are given in the following in the
order of their appearance.

The token {\tt vdecaymode} determines the couplings of the mediator to
the quarks and the DM pair. Our implementation provides the following
two choices:
\begin{itemize}
\item[--] {\tt 1} for pure vector couplings 
\item[--] {\tt 2} for pure axial-vector couplings 
\end{itemize}
The sign of {\tt vdecaymode} is also used to decide whether events
should be generated in the EFT or the full theory: for positive values
the MC simulation is based on the insertion of either ${\cal O}_{V}$
or ${\cal O}_{A}$, while for negative values the program includes the
full $s$-channel propagator of the mediator.

The (mandatory) input parameter {\tt DMmass} (in $\rm GeV$)
specifies the mass $m_\chi$ of the DM particle.

If events are generated in the EFT, the suppression scale $\Lambda$
entering the effective operators should be set with the token {\tt
  DMLambda} (in $\rm GeV$).

If events are generated with the full theory, the user has to provide
values for the mass $M$ of the mediator via {\tt DMVmass} and its
decay width $\Gamma$ via {\tt DMVwidth}. Again units of $\rm GeV$ are
assumed. The couplings of the mediator to the DM particle and the
quarks are, depending on the setting of the parameter {\tt
  vdecaymode}, internally either fixed to $g_\chi^V = g_q^V =1,
g_\chi^A = g_q^A =0$ or $g_\chi^A = g_q^A =1, g_\chi^V = g_q^V =0$.

The character string {\tt runningscale} is used to specify how the renormalisation $\mu_R$ and  factorisation $\mu_F$ scale are determined. The following choices are possible: 
\begin{itemize}
\item[--] {\tt 0} uses fixed scales $\mu_R = \mu_F = 2 \hspace{0.25mm} m_\chi$ 
\item[--] {\tt 1} uses  $\mu_R = \mu_F = p_{T, j_1}$, where $p_{T,j_1}$ is the \\ transverse momentum of the leading jet 
\item[--] {\tt 2} uses $\mu_R = \mu_F = m_{\bar \chi \chi}$, where $m_{\bar \chi \chi}$ \\ denotes the invariant mass of the DM pair
\item[--] {\tt 3} uses $\mu_R = \mu_F = H_T/2$, where $H_T$ is defined in (\ref{eq:HT})
\end{itemize}
In the case of 1, 2 and 3 the scale setting is done
dynamically,~i.e.~the scales are calculated on an event-by-event
basis. If {\tt runningscale} is not specified, the choice 3 is used by
default. This is also the value we used throughout this paper.

The parameter {\tt bornktmin} is used to impose a generation cut on
the minimal $p_T$ of the Born-level process. A suitable choice is
required to achieve an efficient generation of events. To give an
example, if one is interested in events with $p_{T,j_1} > 500 \, {\rm
  GeV}$, a generation cut {\tt bornktmin} of {\tt 450} is a good
choice.

An alternative way of handling this issue is to regulate the Born
divergence by using a suppression factor that damps  the
singularity at $p_T=0$. The functional form implemented in the code
corresponds to $F=p_T^2/(p_T^2 + p_{T,{\rm supp}}^2)$, where
$p_T$ is the transverse momentum of the $\bar \chi \chi$ pair in the
(underlying Born) kinematics.  When the program is run in this mode, a
very small generation cut (e.g.~{\tt bornktmin} $ = 1$) should be
used, and $p_{T,{\rm supp}}$ is set equal to the value of the input
parameter {\tt bornsuppfact}. The program will generate weighted
events, with relative weight given by $1/F$.  Events will be (to a
good approximation) uniformly distributed over the entire $p_T$ range:
the NLOPS accuracy is recovered because, for instance, the few events
at low $p_T$ will have a weight enhanced by $1/F$. In this
way, the correct cross section is reproduced.  More details can be
found in the manuals for $Z/H+ 1$~jet production. If {\tt bornsuppfact} is negative or absent the program runs
with a sharp generation cut, as specified by {\tt bornktmin} (see previous paragraph).

\subsection*{Customising implementation}
\label{sec:modifying}

Our \BOX{} implementation of the jet + $E_{T,{\rm miss}}$ signals is
rather general and the user can take full advantage of this
flexibility by customising the code. Two obvious modifications that
the user might want to implement concern the couplings
$g_{\chi,q}^{V,A}$ and the selection cuts imposed in the analysis.

To modify the couplings of the mediator to the DM particle and the SM
quarks, one has to adjust the code of the three files {\tt Born.f},
{\tt real.f} and {\tt virtual.f}, which are all located in {\tt
  POWHEG-BOX/DMV}. In the file {\tt Born.f} one has to change all the
occurrences of the following code structure\\[3mm]
{\tt
  \noindent
      if(phdm\_mode.eq.'VE') then \\
        \phantom{xxx} ... \\
        \phantom{xxx} jlepZ(:,-3)=Vmm(:) \\
        \phantom{xxx}     jlepZ(:,-1)=Vmp(:) \\
        \phantom{xxx}    jlepZ(:,+1)=Vpm(:) \\
        \phantom{xxx}     jlepZ(:,+3)=Vpp(:) \\
      elseif(phdm\_mode.eq.'AX') then \\
        \phantom{xxx} ... \\
        \phantom{xxx}      jlepZ(:,-3)=Amm(:) \\
        \phantom{xxx}      jlepZ(:,-1)=Amp(:) \\
        \phantom{xxx}      jlepZ(:,+1)=Apm(:) \\
        \phantom{xxx}      jlepZ(:,+3)=App(:) \\
      elseif(phdm\_mode.eq.'bb'.or.phdm\_mode.eq.'ta') then\\
        \phantom{xxx} ...\\
      endif
} \\[3mm]
\noindent The same code structure can be found in {\tt real.f}.
In order to implement~e.g.~a mediator that couples via $\gamma_\mu
\left (1-\gamma_5 \right )$ one would have to change {\tt
  jlepZ(:,-3)=Vmm(:)} into {\tt jlepZ(:,-3)=Vmm(:)-Amm(:)} etc. The
structure of the SM $Z  \bar{b} b$  vertex can be found in the code
contained within the {\tt
  elseif(phdm\_mode.eq.'bb'.or.phdm\_mode.eq.'ta')} statement, and can
be used as a guideline.
In {\tt  virtual.f} the code that sets the mediator couplings takes the form \\[3mm]
{\tt
 \noindent
      if(phdm\_mode.eq.'VE'.or.phdm\_mode.eq.'AX') then\\
      \phantom{xxx} ...\\
      \phantom{xxx} bornamp\_qL(:,:)=bornamp\_qL(:,:)*prop34V \\
      \phantom{xxx} bornamp\_qR(:,:)=bornamp\_qR(:,:)*prop34V \\
      \phantom{xxx} virtamp\_qL(:,:)=virtamp\_qL(:,:)*prop34V \\
      \phantom{xxx} virtamp\_qR(:,:)=virtamp\_qR(:,:)*prop34V \\
      elseif(phdm\_mode.eq.'bb'.or.phdm\_mode.eq.'ta') then\\
        \phantom{xxx} ...\\
      endif \\
      bornamp(-1,:,:)=bornamp\_qL(:,:) \\
      bornamp(+1,:,:)=bornamp\_qR(:,:) \\
      virtamp(-1,:,:)=virtamp\_qL(:,:) \\
      virtamp(+1,:,:)=virtamp\_qR(:,:)
} \\[3mm]
\noindent As above, the $Z \bar b b$ interactions can be found in the
code contained within the statement that begins with {\tt
  elseif(phdm\_mode.eq.'bb'.or.phdm\_mode.eq.'ta')}.

The selection cuts are imposed in the analysis file {\tt
  pwhg\_analysis\_DM\_template.f} that can be found in the directory
{\tt POWHEG-BOX/DMV}. 
The implementation of the cuts is self-explanatory. For instance,~in the case
of the minimum $E_{T,{\rm miss}}$ cut one simply has (after having computed
{\tt Etmiss} from the outgoing momenta)\\[3mm]
{\tt
 \phantom{xxx}           if(Etmiss.lt.min\_Etmiss) goto 666
}  \\[3mm]
\noindent
This means that for events that do not pass the {\tt min\_Etmiss} cut,
the {\tt goto 666} statement is executed,~i.e.~no histogram is
filled. Other cuts can be implemented in a similar fashion.

\end{appendix}

\providecommand{\href}[2]{#2}\begingroup\raggedright\endgroup

\end{document}